\documentclass[aps,prd,nofootinbib,showkeys,preprint,floatfix]{revtex4}

\usepackage{amssymb}
\usepackage{amsmath}
\usepackage{amsfonts}
\usepackage{graphicx}
\usepackage{color}
\usepackage{xspace}
\usepackage{ulem}
\usepackage{lscape}
\usepackage{ wasysym }

\usepackage{multirow,rotating}

\def\gsim{\raise0.3ex\hbox{$\;>$\kern-0.75em\raise-1.1ex\hbox{$\sim\;$}}}
\def\lsim{\raise0.3ex\hbox{$\;<$\kern-0.75em\raise-1.1ex\hbox{$\sim\;$}}}

\def\pslash{p \hspace{-0.5em}/\;\:}

\def\znbb{0\nu\beta\beta}
\def\meff{\langle m_{\nu} \rangle}

\newcommand{\ba}[1]{\begin{eqnarray} \label{(#1)}}
\newcommand{\ea}{\end{eqnarray}}

\newcommand{\AddrAHEP}{
  {\it AHEP Group, Instituto de F\'{\i}sica Corpuscular --
    C.S.I.C./Universitat de Val{\`e}ncia \\
    Edificio Institutos de Investigacion, Parc Cientific de Paterna, 
  Apartado 22085,
  E--46071 Val{\`e}ncia, Spain}}

\newcommand{\AddrUFSM}{
Universidad T\'ecnica Federico Santa Mar\'\i a, \\ 
Centro-Cient\'\i fico-Tecnol\'{o}gico de Valpara\'\i so, \\ 
Casilla 110-V, Valpara\'\i so,  Chile}

\newcommand{\AddrSaitama}{
Department of Physics, Saitama University, \\ 
Shimo-Okubo 255, 338-8570 Saitama-Sakura, Japan}

\def\gsim{\raise0.3ex\hbox{$\;>$\kern-0.75em\raise-1.1ex\hbox{$\sim\;$}}}
\def\lsim{\raise0.3ex\hbox{$\;<$\kern-0.75em\raise-1.1ex\hbox{$\sim\;$}}}

\allowdisplaybreaks[4]

\begin{document}

\preprint{IFIC/15-82, STUPP-15-224}

\title{Long-range contributions to double beta decay revisited}

\author{J.C. Helo} \email{juancarlos.helo@usm.cl}\affiliation{\AddrUFSM}
\author{M. Hirsch} \email{mahirsch@ific.uv.es}\affiliation{\AddrAHEP}
\author{T. Ota}\email{toshi@mail.saitama-u.ac.jp}\affiliation{\AddrSaitama}



\begin{abstract}

We discuss the systematic decomposition of all dimension-7 ($d=7$)
lepton number violating operators.  These $d=7$ operators produce
momentum enhanced contributions to the long-range part of the $\znbb$
decay amplitude and thus are severely constrained by existing
half-live limits. In our list of possible models one can find
contributions to the long-range amplitude discussed previously in the
literature, such as the left-right symmetric model or scalar
leptoquarks, as well as some new models not considered before.  The
$d=7$ operators generate Majorana neutrino mass terms either at
tree-level, 1-loop or 2-loop level. We systematically compare
constraints derived from the mass mechanism to those derived from the
long-range $\znbb$ decay amplitude and classify our list of models
accordingly. We also study one particular example decomposition, which
produces neutrino masses at 2-loop level, can fit oscillation data and
yields a large contribution to the long-range $\znbb$ decay amplitude,
in some detail.

\end{abstract}

\maketitle

\tableofcontents


\section{Introduction}

Majorana neutrino masses, lepton number violation and neutrinoless
double beta decay ($\znbb$) are intimately related. It is therefore
not surprising that many models contributing to $\znbb$ have been
discussed in the literature, see for example the recent reviews
\cite{Deppisch:2012nb,Hirsch:2015cga}. However, the famous black-box
theorem \cite{Schechter:1981bd} guarantees only that - if $\znbb$ decay
is observed - Majorana neutrino masses must appear at the 4-loop level,
which is much too small \cite{Duerr:2011zd} to explain current
oscillation data \cite{Forero:2014bxa}.  Thus, a priori one does not
know whether some ``exotic'' contribution or the mass mechanism
dominates the $\znbb$ decay rate. Distinguishing the different
contributions would not only be an important step towards determining
the origin of neutrino masses, but would also have profound
implications for leptogenesis
\cite{Deppisch:2013jxa,Deppisch:2015yqa}.

In terms of only standard model (SM) fields, $\Delta L=2$ terms can be
written as non-renormalizable operators (NROs) of odd mass dimensions.
At mass dimension $d=5$, there is only one such operator, the famous
Weinberg operator \cite{Weinberg:1979sa}, ${\cal O}_W =
\frac{1}{\Lambda}(LLHH)$.  At tree-level the Weinberg operator can be
understood as the low-energy limit of one of the three possible seesaw
realizations
\cite{Yanagida:1979as,GellMann:1980vs,Foot:1988aq,Mohapatra:1979ia,Ma:1998dn}.
All other $\Delta L=2$ operators up to $d=11$ --- excluding, however,
possible operators containing derivatives --- have been listed in
\cite{Babu:2001ex}.  When complemented with SM Yukawa interactions
(and in some cases SM charged current interactions), these higher
dimensional operators always also generate Majorana neutrino masses
(at different loop-levels), leading again to the Weinberg
operator\footnote{Or to operators of the form ${\cal O}_W\times
  (H^{\dagger}H)^n$, $n=1,2,\cdots$. For neutrino mass models based on
  this type of effective operators, see
  e.g.,~\cite{Giudice:2008uua,Babu:2009aq,Bonnet:2009ej,Picek:2009is,Kanemura:2010bq,Krauss:2011ur,Krauss:2013gy,Bambhaniya:2013yca}.}
  at low energies.

All $\Delta L=2$ operators also contribute to $\znbb$ decay.  From the
nuclear point of view, the amplitude for $\znbb$ decay contains
two parts: the long-range part and the short-range part.
The so-called long-range part \cite{Pas:1999fc} describes
all contributions involving the exchange of a light, virtual neutrino
between two nucleons. This category contains the mass mechanism, 
i.e. the Weinberg operator sandwiched between two SM charged current
interactions, and also contributions due to $d=7$ lepton number
violating operators.\footnote{%
We save the term ``long-range contribution'' for the contribution from
the $d=7$ operators and call the standard contribution from 
Majorana neutrino mass separately the ``mass mechanism''. }
The short-range part of the $\znbb$ decay amplitude \cite{Pas:2000vn},
on the other hand, contains all contributions from the exchange of
heavy particles and can be described by a certain subset of the $d=9$
$\Delta L=2$ operators in the list of \cite{Babu:2001ex}.  In total
there are six $d=9$ operators contributing to the short-range part of
the amplitude at tree-level and the complete decomposition for the
(scalar induced) operators has been given in \cite{Bonnet:2012kh}. The
relation of all these decompositions with neutrino mass models has
been studied recently in \cite{Helo:2015fba}.\footnote{%
Neutrino mass models based on the $\Delta L =2$ effective
operators were discussed in \cite{Babu:2001ex,deGouvea:2007xp}  
The decomposition of the $\Delta L=2$ operators was also discussed in
  \cite{Angel:2012ug,Angel:2013hla}.}  
The general conclusion of \cite{Helo:2015fba} is that for 2-loop and
3-loop neutrino mass models, the short-range part of the amplitude
could be as important as the mass mechanism, while for tree-level and
1-loop models one expects that the mass mechanism gives the dominant
contribution to $\znbb$ decay.\footnote{%
Possible LHC constraints 
on short-range operators contributing to $\znbb$ decay have 
been discussed in \cite{Helo:2013ika,Helo:2015ffa}.}

In this paper we study $d=7$ $\Delta L=2$ operators, their relation to
neutrino masses and the long-range part of the $\znbb$ decay
amplitude. We decompose all $d=7$ $\Delta L=2$ operators and determine
the level of perturbation theory, at which the different
decompositions (or ``proto-models'') will generate neutrino masses.
Tree-level, 1-loop and 2-loop neutrino mass models are found in the
list of the decompositions.  We then compare the contribution from the
mass mechanism to the $\znbb$ decay amplitude with the long-range
$d=7$ contribution.  Depending on which particular {\em nuclear}
operator is generated, limits on the new physics scale $\Lambda \gsim
g_{\rm eff} (17-180)$ TeV can be derived from the $d=7$ contribution.
Here, $g_{\rm eff}$ is the mean of the couplings entering the (decomposed)
$d=7$ operator.  This should be compared to limits of the order of
roughly $\Lambda \gsim \sqrt{Y_{\rm eff}}$ $10^{11}$ TeV and $\Lambda
\gsim Y_{\rm eff}^2$ $50$ TeV, derived from the upper limit on $\meff$
for tree-level and 2-loop ($d=7$) neutrino masses. (Here, $Y_{\rm
  eff}$ is again some mean of couplings entering the neutrino mass
diagram. We use a different symbol, to remind that $Y_{\rm eff}$ is
not necessarily the same combination of couplings as $g_{\rm eff}$.)
Thus, only for a certain, well-defined subset of models can the
contribution from the long-range amplitude be expected to be similar
to or dominate over the mass mechanism. Note that, conversely a
sub-dominant contribution to the long-range amplitude always exists
also in all models with mass mechanism dominance.

We then give the complete classification of all models contributing to
the $d=7$ operators in tabular form in the appendix of this paper.  In
this list all models giving long-range contributions to $\znbb$ decay
can be found, such as, for example, supersymmetric models with
R-parity violation \cite{Babu:1995vh,Pas:1998nn} or scalar leptoquarks
\cite{Hirsch:1996ye}. There are also models with non-SM vectors, which
could fit into models with extended gauge sectors, such as the
left-right symmetric model
\cite{Pati:1974yy,Mohapatra:1974gc,Mohapatra:1980yp}. And, finally, 
there are new models in this list, not considered in the literature 
previously.

We mention that our paper has some overlap with the recent
work~\cite{Cai:2014kra}.  The authors of this paper also studied $d=7$
$\Delta L=2$ operators.\footnote{%
Decompositions of $d=7$ operators were also discussed in
  \cite{Lehman:2014jma,Bhattacharya:2015vja}.}  They discuss 1-loop
neutrino masses induced by these operators, lepton flavour violating
decays and, in particular, LHC phenomenology for one example operator
in detail. The main differences between our work and theirs is that we
(a) focus here on the relation of these operators with the long-range
amplitude of $\znbb$ decay, which was not studied in
\cite{Cai:2014kra} and (b) also discuss tree-level and 2-loop neutrino
mass models.  In particular, we find that 2-loop neutrino mass models
are particularly interesting, because the $d=7$ long-range
contribution dominates $\znbb$ only in the class of models.

The rest of this paper is organized as follows. In the next section we
lay the basis for the discussion, establishing the notation and
recalling the main definitions for $\Delta L=2$ operators and $\znbb$
decay amplitude. In the following section we then discuss an example
of each: tree-level, 1-loop and 2-loop neutrino mass models. In each
case we estimate the contribution to the mass mechanism and the
constraints from the long-range amplitude. We study a 2-loop $d=7$
model in some more detail, comparing also to oscillation data
and discuss the constraint from lepton flavour violating processes. 
In section \ref{sect:lim} we then discuss a special case, where a $d=9$
operator can give an equally important contribution to the $\znbb$
decay amplitude as a $d=7$ operator. The example we discuss is related
to the left-right symmetric extension of the standard model and, thus, 
of particular interest. We then close the paper with a short summary. The
complete list of decompositions for $d=7$ operators is given as an
appendix.

\section{General setup}
\label{sect:setup}

The $\znbb$ decay amplitude can be separated into two pieces: (a) the
long-range part \cite{Pas:1999fc}, including the well-known mass
mechanism, and (b) the short-range part \cite{Pas:2000vn} of the decay
rate describing heavy particle exchange. Here, we will concentrate
exclusively on the long-range part of the amplitude.

The long-range part of the amplitude exchanges a light, virtual
neutrino between two point-like vertices. The numerator of the
neutrino propagator involves two pieces, $(m_{\nu_i} + \pslash)$.  If
the interaction vertices contain standard model charged current
interactions, the $m_{\nu_i}$-term is projected out. This yields the
``mass mechanism'' of $\znbb$ decay. However, if one of the two
vertices involved in the diagram produces a neutrino in the wrong
helicity state, i.e. $(\nu_L)^c$, the $\pslash$-term is picked from
the propagator. Since the momentum of the virtual neutrino is
typically of the order of the Fermi momentum of the nucleons, $p_F
\simeq 100$ MeV, the $\znbb$ amplitude from the operators proportional
to $\pslash$ is enhanced by $p_{F}/m_{\nu} \gtrsim
\mathcal{O}(10^{8})$ with respect to the amplitude of the standard
mass mechanism.  Consequently, any operator proportional to $\pslash$
will be tightly constrained from non-observation of double beta decay.
Following \cite{Pas:1999fc} we write the effective Lagrangian for
4-fermion interactions as
\begin{eqnarray}\nonumber
\mathcal{L}^{\text{4-Fermi}}
&=& \mathcal{L^{\rm SM}} + \mathcal{L^{\rm LNV}} \\  
           & =& \frac{G_F}{\sqrt{2}}
           \left[
           j^{\mu}_{V-A} J_{V-A,\mu}
           +
            \hspace{-0.3cm}
           \sum_{\begin{minipage}{1.5cm}
                    {\tiny 
                      $\alpha,\beta \neq V-A$ 
                    }
                  \end{minipage}} 
            \hspace{-0.3cm}
            \epsilon_{\alpha}^{\beta} 
            \hspace{0.1cm}
            j_{\beta}J_{\alpha}\right]\,.\label{eq:defLR}
\end{eqnarray}
The leptonic (hadronic) currents $j_{\beta}$ ($J_{\alpha}$) 
are defined as:
\begin{gather}\label{eq:CurrLR}
J^{\mu}_{V\pm A}
=
(J_{R/L})^{\mu}
\equiv
\overline{u}\gamma^{\mu}(1\pm\gamma_5)d\,,
\qquad
j_{V\pm A}^{\mu}
\equiv
\overline{e}\gamma^{\mu}(1\pm\gamma_5)\nu\,,
\\ \nonumber
J_{S\pm P}
=
J_{R/L}
\equiv
\overline{u}(1\pm\gamma_5)d\,,
\qquad
j_{S\pm P}\equiv\overline{e}(1\pm\gamma_5)\nu\,,
\\ \nonumber
J^{\mu\nu}_{T_{R/L}}
=
(J_{R/L})^{\mu \nu}
\equiv
\overline{u}\gamma^{\mu\nu}(1\pm\gamma_5)d\,,
\qquad
j_{T_{R/L}}^{\mu\nu} \equiv\overline{e}\gamma^{\mu\nu}(1\pm\gamma_5)\nu\ ,
\end{gather}
where $\gamma^{\mu \nu}$ is defined as $\gamma^{\mu \nu} = \frac{\rm
  i}{2} [\gamma^{\mu}, \gamma^{\nu}]$.  The first term of
Eq.~\eqref{eq:defLR} is the SM charged current interaction, the other
terms contain all new physics contributions. We normalize the
coefficients $\epsilon_{\alpha}^{\beta}$ relative to the SM charged
current strength $G_F/\sqrt{2}$. Recall, $P_{L/R} =\frac{1}{2}(1 \mp
\gamma_5)$ and we will use the subscripts $L$ and $R$ for left-handed
and right-handed fermions, respectively. Note also that all leptonic
currents with $(1 -\gamma_5)$ will pick $m_{\nu_i}$ from the
propagator, leading to an amplitude proportional to
$\epsilon^{\beta_L}_{\alpha}\times \meff$ ($\beta_{L} \in \{S-P, V-A,
T_{L}\}$), which is always smaller than the standard mass mechanism
contribution and thus is not very interesting.  Thus, only six
particular $\epsilon^{\beta}_{\alpha}$ can be constrained from $\znbb$
decay. For convenience, we repeat the currently best limits, all
derived in \cite{Deppisch:2012nb}, in Table \ref{Tab:LimitsLong}.

\begin{table}[t]
\centering
\begin{tabular}{ccccccc}
\hline
Isotope & $|\epsilon^{V+A}_{V-A}|$ &  $|\epsilon^{V+A}_{V+A}|$ & 
          $|\epsilon^{S+P}_{S-P}|$ &  $|\epsilon^{S+P}_{S+P}|$ & 
          $|\epsilon^{TR}_{TL}|$   &  $|\epsilon^{TR}_{TR}|$   \\
\hline
$^{136}$Xe & $2.0 \cdot 10^{-9}$  & $3.9 \cdot 10^{-7}$   & 
          $4.7 \cdot 10^{-9}$    & $4.7 \cdot 10^{-9}$   & 
          $3.3 \cdot 10^{-10}$   & $5.6\cdot 10^{-10}$    \\
\hline
\end{tabular}
\caption{Limits on $\epsilon_{\alpha}^{\beta_{R}}$ from
  non-observation of $^{136}$Xe $\znbb$ decay, where $\beta_{R} \in \{S+P,
 V+A, T_{R}\}$.  These limits were
  derived in \cite{Deppisch:2012nb} and have been updated with the
  combined limit from KamLAND-Zen and Exo-200 \cite{Gando:2012zm}.} 
\label{Tab:LimitsLong}
\end{table}

Eq.(\ref{eq:defLR}) describes long-range $\znbb$ decay from the
low-energy point of view. From the particle physics point of view,
these $\Delta L=2$ currents can be described as being generated from
$d=7$ operators.  Disregarding the $d=7$ ``Weinberg-like'' operator
${\cal O}_W\times (H^{\dagger}H)$, there are four of these operators
in the list of Babu \& Leung \cite{Babu:2001ex}:
\begin{eqnarray}\label{eq:BL}
{\cal O}_2 &\propto& L^iL^jL^ke^cH^l \epsilon_{ij}\epsilon_{kl} ,
\\ \nonumber
{\cal O}_3 \equiv \{ {\cal O}_{3a} ,  {\cal O}_{3b} \} 
          & \propto & \{ L^iL^jQ^kd^cH^l \epsilon_{ij}\epsilon_{kl} ,
                         L^iL^jQ^kd^cH^l \epsilon_{ik}\epsilon_{jl} \} ,
\\ \nonumber
{\cal O}_4 \equiv \{ {\cal O}_{4a} ,  {\cal O}_{4b} \} 
          & \propto & \{ L^iL^j{\bar Q}_i{\bar u^c}H^k \epsilon_{jk},
                         L^iL^j{\bar Q}_k{\bar u^c}H^k \epsilon_{ij} \} ,
\\ \nonumber
{\cal O}_8 &\propto& L^i{\bar e^c}{\bar u^c}d^cH^j \epsilon_{ij} .
\end{eqnarray}
Here, ${\cal O}_2$ is included for completeness, although it is
trivial that the mass mechanism will be the dominant contribution to
$\znbb$ decay for this operator, since it does not involve any quark
fields. We will therefore not discuss the detailed decomposition of
${\cal O}_2$, which can be found in \cite{Cai:2014kra}. The operators
${\cal O}_{3b,4a,8}$ will contribute to the long-range amplitudes
$j_{\beta}J_{\alpha}$, and the coefficient of the amplitudes is
described as
\begin{equation}\label{eq:defe7}
\frac{G_F \epsilon_{d=7}}{\sqrt{2}}\simeq \frac{g_{\rm eff}^3 v}{4\Lambda_7^3},
\end{equation}
where $\Lambda_{7}$ is the energy scale from which the $d=7$ operators
originate, and $\epsilon_{d=7}$ is one of (or a combination of two of)
the $\epsilon^{\beta}_\alpha$ of Table 1. The factor $1/4$ is included
to account for the fact that Eq.~(\ref{eq:CurrLR}) is written in terms
of $(1\pm \gamma_5$) while chiral fields are defined using
$P_{L/R}$. This leads to the numerical constraints on the scale
$\Lambda_7$ mentioned in the introduction, taking the least/most
stringent numbers from Table \ref{Tab:LimitsLong}.

All $\Delta L=2$ operators generate Majorana neutrino masses. However,
operators ${\cal O}_{3a}$ and ${\cal O}_{4b}$ will generate 
neutrino mass matrices without diagonal entries, 
since $L^iL^j\epsilon_{ij}=0$ within a generation.  
Neutrino mass matrices with such a flavour structure
result in very restricted
neutrino spectra, and it was shown in \cite{Koide:2001xy} that such
models necessarily predict $\sin^2(2\theta_{12}) = 1 - (1/16)
(\Delta m^2_{21}/\Delta m^2_{31})^2$. This prediction is ruled
out by current neutrino data at more than 8 $\sigma$ c.l.
\cite{Forero:2014bxa}. Models that generate at low energies only
${\cal O}_{3a}$ or ${\cal O}_{4b}$ can therefore not be considered
realistic explanation of neutrino data.\footnote{%
However, models that produce these operators usually 
allow to add additional interactions that will generate ${\cal O}_{5}$ 
(${\cal O}_{6}$) in addition to  ${\cal O}_{3a}$ (${\cal O}_{4b}$),  
as for example in the model discussed in \cite{Babu:2011vb}. These 
constructions then allow to correctly explain neutrino oscillation 
data, since  ${\cal O}_{5}$/${\cal O}_{6}$ produce non-zero elements 
in the diagonal entries of the neutrino mass matrix.}

Flavour off-diagonality of ${\cal O}_{3a}$ and ${\cal O}_{4b}$ does
also suppress strongly their contribution to long-range double beta
decay, in case the resulting leptonic current is of type $j_{S+P}$
(see appendix\footnote{%
Decomposition \#8 of $\mathcal{O}_{3a}$ also
generates $j_{T_{R}}$ which can contribute to $\znbb$ without the
need for a non-unitarity of the mixing matrix.}).  This is because
the final state leptons are both electrons, while the virtual neutrino
emitted from the $L$ in ${\cal O}_{3a,4b}$ is necessarily either
$\nu_{\mu}$ or $\nu_{\tau}$. In the definition of the ``effective''
$\epsilon_{\alpha}^{\beta}$, then neutrino mixing matrices appear with
the combination $\sum_j U_{ej} U_{\mu j}^{*}$ (or $U_{ej}U_{\tau j}^{*}$), 
which is identically zero unless the mixing matrices are
non-unitary when summed over the light neutrinos. 

Departures from unitarity can occur in models with extra
(sterile/right-handed) neutrinos heavier than about $\sim 1$ GeV.
While the propagation of the heavy neutrinos also contributes to
$\znbb$, the nuclear matrix element appearing in the amplitude of the
heavy neutrino exchange is strongly suppressed, when their masses are
larger than 1 GeV~\cite{Haxton:1985am,Muto:1989cd}. 
Consequently, the heavy neutrino contribution is
suppressed with respect to the light neutrino one and the sum over
$\sum_j U_{ej} U_{\mu j}^{*}$ is incomplete, appearing effectively as
a sum over mixing matrix elements which is non-unitary.  Current
limits on this non-unitary piece of the mixing are of the order of
very roughly percent
\cite{Antusch:2006vwa,Akhmedov:2013hec,Antusch:2014woa,Escrihuela:2015wra},
thus weakening limits on the coefficients for ${\cal O}_{3a}$ and
${\cal O}_{4b}$ (for $j_{S+P}$), compared to other operators, by {\em
  at least} two orders of magnitude.

To the list in Eq.~(\ref{eq:BL}) one can add two more $\Delta L=2$ 
operators involving derivatives:
\begin{eqnarray}\label{eq:d7D}
{\cal O}_1^{D_{\mu}} \equiv \{ {\cal O}_{1a}^{D_{\mu}} ,  {\cal O}_{1b}^{D_{\mu}} \} 
& \propto &  \{ L^iL^jD_{\mu}D_{\mu}H^kH^l  \epsilon_{ij}\epsilon_{kl} ,
           L^iL^jD_{\mu}D_{\mu}H^kH^l  \epsilon_{ik}\epsilon_{jl}  \} 
\\ \nonumber
{\cal O}_2^{D_{\mu}}& \propto & L^ie^cD_{\mu}H^jH^kH^l \epsilon_{ij}\epsilon_{kl}
\end{eqnarray}
We mention these operators for completeness.
As shown in \cite{delAguila:2012nu}, 
tree-level decompositions of ${\cal O}_{1}^{D_{\mu}}$ 
always involve one of the seesaw mediators,
and thus one expects this operator to be always
present in tree-level models of neutrino mass. 
As we will see, 
if neutrino masses are generated from tree-level, 
the mass mechanism contribution in general dominates $\znbb$,
and consequently 
the new physics effect from ${\cal O}_{1}^{D_{\mu}}$ 
cannot make a measurable impact.
The second type of the derivative operators, 
${\cal O}_2^{D_{\mu}}$, 
has also been discussed in detail in \cite{delAguila:2012nu}
with an example of tree-level realization, 
we thus give only a brief summary for this operator 
in the appendix.

\section{Classification}
\label{sect:class}

In this section we will discuss a classification scheme for the
decompositions of the $\Delta L=2$ operators of Eq.~(\ref{eq:BL}),
based on the number of loops, at which they generate neutrino masses.
We will discuss one typical example each for tree-level, 1-loop and
2-loop models. The complete list of decompositions for the different 
cases can be found in the appendix.

\subsection{Tree level}
\label{subsect:tree}

If the neutrino mass is generated at tree-level, one expects $m_{\nu}
\propto v^2/\Lambda$, which for coefficients of ${\cal O}(1)$ give
$\Lambda \sim 10^{14}$ GeV for neutrino masses order $0.1$ eV. The
amplitude of the mass mechanism of $\znbb$ decay is proportional to
${\cal A}^{\rm MM} \propto \meff/p_F^2 \times (1/m_W^2)^2$, while the
amplitude provided from the $d=7$ operator is ${\cal A}^{\rm LR} \propto p_F
v/(\Lambda^3p_F^2) \times (1/m_W^2)$. The $d=7$ contribution is therefore
favoured by a factor $p_F/\meff$, but suppressed by $(v/\Lambda)^3$.
Inserting $\Lambda \sim 10^{14}$, the $d=7$ amplitude
should be smaller than the mass mechanism amplitude 
by a huge factor of order ${\cal O}(10^{-27})$. 
However, this naive estimate assumes all
coefficients in the operators to be order ${\cal O}(1)$. Since these
coefficients are usually products of Yukawa (and other) couplings in
the UV complete models, this is not necessarily the case in general
and much smaller scales $\Lambda$ could occur.

\begin{figure}[t]
\hskip-10mm\includegraphics[width=0.45\linewidth]{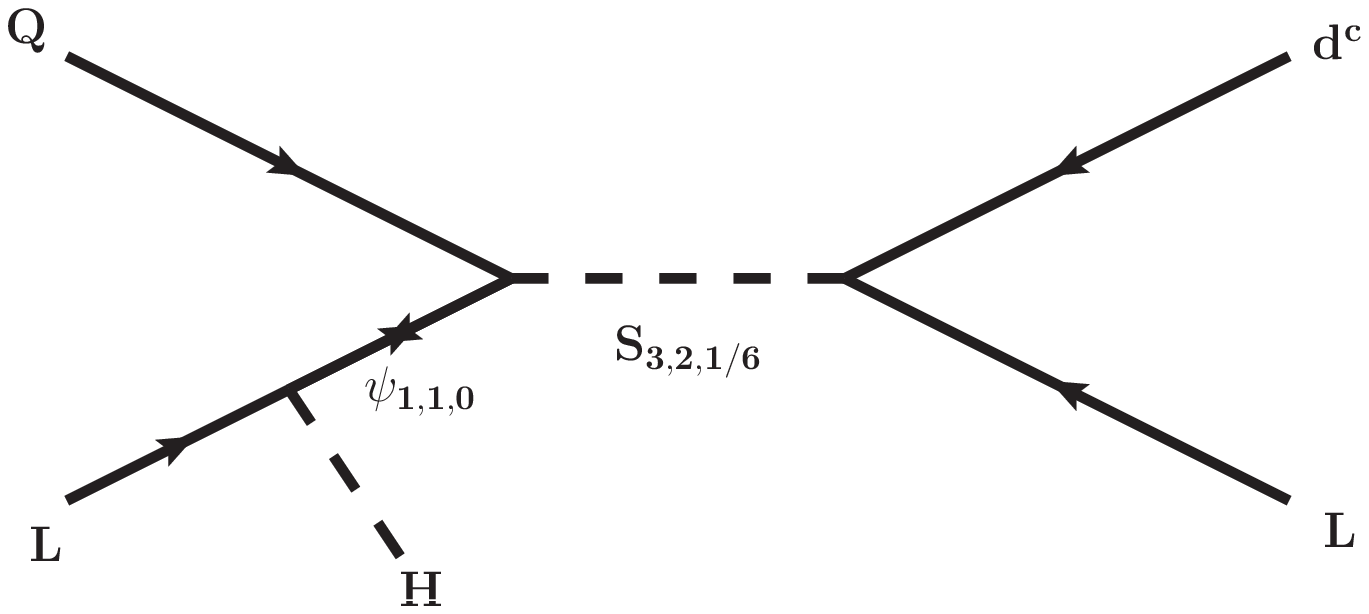}
\hskip10mm\includegraphics[width=0.45\linewidth]{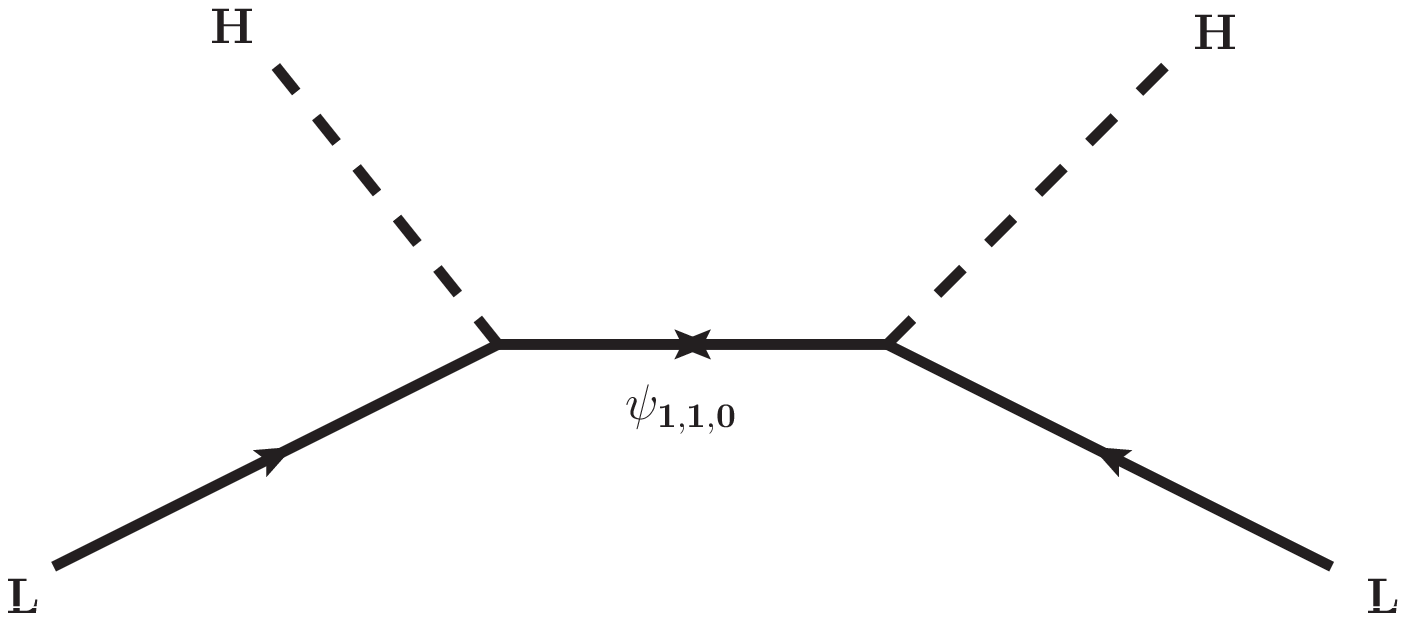}
\caption{To the left: Diagram leading to long-range $\znbb$ decay via
  charged scalar exchange for Babu-Leung operator ${\cal O}_{3}$
  (BL\#3). To the right: Tree-level neutrino mass generated via seesaw
  type-I, using the same vertices as in the diagram on the left. Here
  and in all Feynman diagrams below, arrows on fermion lines indicate
  the flow of particle number, not the chirality of the fermion. The 
  double arrow on $\psi_{1,1,0}$ indicates its Majorana nature.}
\label{fig:SSI}
\end{figure}

To discuss this in a bit more detail, we consider a particular example
based on ${\cal O}_3$, decomposition $\#$4, where two new fields, (1)
a Majorana fermion $\psi$ with the SM charge $(SU(3)_{c}, SU(2)_{L},
U(1)_{Y})=(1,1,0)$ and (2) a scalar $S$ with $(3,2,1/6)$, are
introduced to decompose the effective operator, see Table
\ref{Tab:BL3} and Fig.~\ref{fig:SSI}.  The Lagrangian for this model
contains the following terms:
\begin{eqnarray}\label{eq:o3d4}
{\cal L}_{\text{3,\#4}} &=&
 Y_{\nu} {\overline \psi} H \cdot L
 + 
 Y_{d^c L} \overline{d_{R}} L \cdot S 
 + 
 Y_{Q\psi} {\overline\psi} Q \cdot S^{\dagger}
 + 
 m_{\psi} {\overline{\psi^{c}}} \psi 
 + 
 {\rm h.c}.
\end{eqnarray}
Here, we have suppressed generation indices for simplicity.  The first
term in Eq.~(\ref{eq:o3d4}) will generate Dirac masses for the
neutrinos.  The Majorana mass term for the neutral field $\psi$
(equivalent to a right-handed neutrino) can not be forbidden in this
model. We will discuss first the simplest case with only one copy of
$\psi$ and comment on the more complicated cases with two or three
$\psi$ below.

The contribution to $\znbb$ decay can be read off directly from 
the diagram in Fig.~\ref{fig:SSI} on the left. It is given by 
\begin{eqnarray}\label{eq:eps34}
\frac{G_F \epsilon_{{\cal O}_{3,\#4}}}{\sqrt{2}}\simeq 
\frac{(Y_{\nu})_e v}{m_{\psi}}
\frac{ (Y_{d^cL})_{1e}(Y_{Q\psi})_1}{m_{S}^2}.
\end{eqnarray}
With only one copy of $\psi$, the effective mass term contributing 
to $\znbb$ decay is $\meff = (Y_{\nu})_e^2v^2/m_{\psi}$ 
and we can replace $(Y_{\nu})_e$ by $\meff$ to arrive at the rough 
estimate of the constraint derived from the 
$d=7$ contribution to $\znbb$:
\begin{equation}\label{eq:limO3}
(Y_{d^cL})_{1e}(Y_{Q\psi})_1 \lsim 3 \times 10^{-3}
  \left( \frac{\meff}{\rm 0.5 \hskip1mm eV} \right)^{-1/2}
  \left( \frac{m_{\psi}}{\rm 100 \hskip1mm GeV} \right)^{1/2}
  \left( \frac{m_{S}}{\rm 1 \hskip1mm TeV} \right)^{2}
\end{equation}
Eq.~(\ref{eq:limO3}) shows that the upper limit on the Yukawa
couplings disappears as $\meff$ approaches zero.  When the masses are
greater than roughly $m_{\psi} \simeq m_{S} \sim 10$ TeV, the Yukawa
couplings must be non-perturbative to fulfil the equality in
Eq.~\eqref{eq:limO3}.  This implies that the mass mechanism will
always dominate the $\znbb$ contribution for scales $\Lambda$ larger
than roughly this value, independent of the exact choice of the
couplings.

We briefly comment on models with more than one $\psi$.  As is
well-known, neutrino oscillation data require at least two non-zero
neutrino masses, while a model with only one $\psi$ leaves two of the
three active neutrinos massless. Any realistic model based on
Eq.~(\ref{eq:o3d4}) will therefore need at least two copies of
$\psi$. In this case Eq.~(\ref{eq:eps34}) has to be modified to
include the summation over the different $\psi_{i}$ and
$\epsilon_{{\cal O}_{3,\#4}} \propto \sum_i\frac{(Y_{\nu})_{ei} v}
{m_{\psi_i}}$. $\meff$, on the other hand, is proportional to $\meff
\propto \sum_i \frac{(Y_{\nu})_{ei}^2}{m_{\psi_i}}$. In this case, one
still expects in general that limits derived from the long-range part
of the amplitude are proportional to $\meff$. However, there is a
special region in parameter space, where the different contributions
to $\meff$ cancel nearly exactly, leaving the long-range contribution
being the dominant part of the amplitude.  Unless the model parameters
are fine-tuned in this way, the mass mechanism should win over the
$d=7$ contribution for all tree-level neutrino mass models.

The tables in the appendix show, that all three types of seesaw
mediators appear in the decompositions of ${\cal O}_3$, ${\cal O}_4$
and ${\cal O}_8$: $\psi_{1,1,0}$ (type-I), $\psi_{1,3,0}$ (type-III)
and $S_{1,3,1}$ (type-II). In order to generate a seesaw mechanism,
for some of the decompositions one needs to introduce new
interactions, such as $S_{1,3,1}^{\dagger}HH$, not present in the
corresponding decomposition itself. However, in all these cases, the
additional interactions are allowed by the symmetries of the models
and are thus expected to be present. One then expects for all
tree-level decompositions that the mass mechanism dominates over the
long-range part of the amplitude, unless (i) the new physics scale
$\Lambda$ is below a few TeV {\em and} (ii) some parameters are
extremely fine-tuned to suppress light neutrino masses, as discussed
above in our particular example decomposition.

\subsection{One-loop level}
\label{subsect:1lp}

\begin{figure}[t]
\begin{center}
\unitlength=1cm
 \begin{picture}(16,5)
\put(0.5,0){\includegraphics[width=0.35\linewidth]{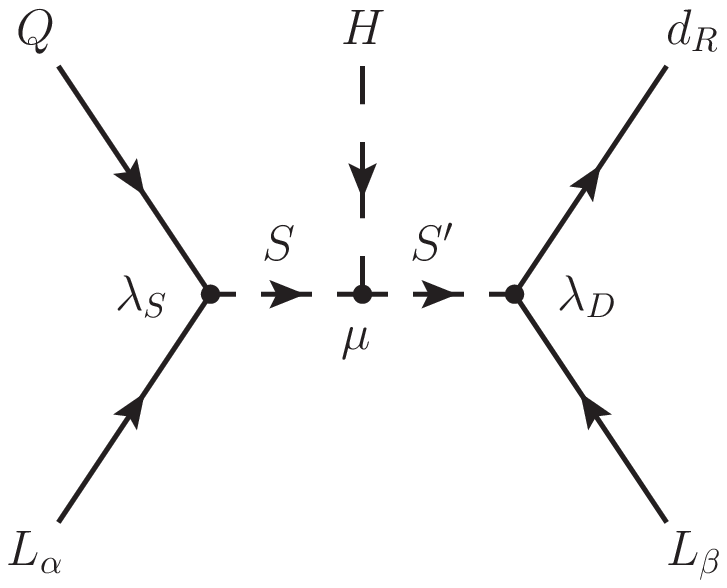}}
\put(8.5,0.3){\includegraphics[width=0.4\linewidth]{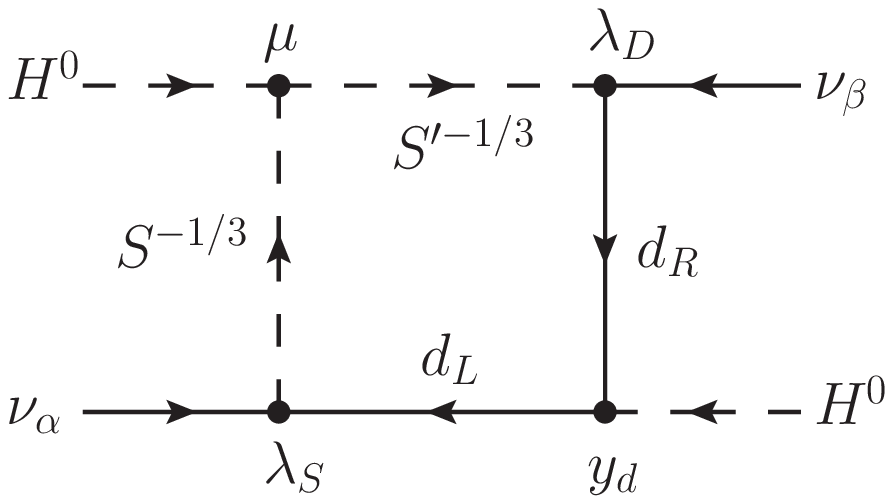}}
 \end{picture}
\caption{Decomposition \#2 of ${\cal O}_{3}$ operator (left)
 and one-loop diagram for neutrino masses based on the 
 decomponsition (right).}
\label{fig:1dim7}
\end{center}
\end{figure}

We now turn to a discussion of one-loop neutrino mass models. 
For this class of neutrino mass models, naive estimates 
would put $\Lambda$ at $\Lambda \sim {\cal O}(10^{12})$ GeV 
for coefficients of ${\cal O}(1)$ and neutrino masses of 
$\mathcal{O}(0.1)$ eV. 
Thus, in the same way as tree neutrino mass models,
the mass mechanism dominates over the long-range amplitude, 
unless at least some of the couplings in the UV completion
are significantly smaller than ${\cal O}(1)$, as discussed next.

As shown in \cite{Bonnet:2012kz}, there are only three genuine 1-loop
topologies for ($d=5$) neutrino masses.  Decompositions of 
${\cal O}_3$, ${\cal O}_4$ or ${\cal O}_8$ produce only two of them,
namely T$\nu$-I-ii or T$\nu$-I-iii.  We will discuss one example for
T$\nu$-I-ii, based on ${\cal O}_3$ decomposition $\# 2$, see 
Table \ref{Tab:BL3} and Fig.~\ref{fig:1dim7}.
The underlying leptoquark model was first discussed
in \cite{Hirsch:1996qy,Hirsch:1996ye}, and for accelerator phenomenology
see, e.g., \cite{AristizabalSierra:2007nf}. 
The model adds two scalar states to the SM particle content, 
$S (3,1,-1/3)$ and $S' (3,2,1/6)$. 
The Lagrangian of the model contains interactions
with SM fermions
\begin{equation}\label{eq:lagLQ}
{\cal L}_{3,\#2}^{\text{LQ}} 
 = 
 (\lambda_S)_{\alpha i} 
 {\overline{L_{\alpha}^c}} \cdot Q_{i} S^\dagger 
 + (\lambda_D)_{i \alpha} 
 {\overline{d_{R i}}} L_{\alpha} \cdot S'
 + \cdots,
\end{equation}
and the scalar interactions and mass terms:
\begin{equation}\label{eq:lagLQH}
{\cal L}_{3,\#2}^{\text{scalar}} 
 = 
 \mu 
 S H \cdot S'^\dagger 
 +
 m_S^2 |S|^2 
 +
 m_D^2 |S'|^2 
 + \cdots
\end{equation}
Lepton number is violated by the simultaneous presence of the terms in
Eq.~(\ref{eq:lagLQ}) and the first term in 
Eq.~(\ref{eq:lagLQH})~\cite{Hirsch:1996qy}.  
Electro-weak symmetry breaking generates 
the off-diagonal element of the mass
matrix for the scalars with the electric charge $-1/3$. 
The mass matrix is expressed as
\begin{equation}\label{eq:massLQ}
M^2_{\rm LQ} =
\begin{pmatrix}
m_S^2 & \mu v \\
 \mu v & m_D^2
\end{pmatrix}
\end{equation}
in the basis of ($S^{-1/3}, S'^{-1/3}$),
which is diagonalized by the rotation matrix with the 
mixing angle $\theta_{\rm LQ}$ that is given as
\begin{equation}\label{eq:AngLQ}
\tan2\theta_{\rm LQ} = \frac{2 \mu v}{m_S^2 -m_D^2}.
\end{equation}
The neutrino mass matrix, 
which arises from the 1-loop diagram 
shown in Fig.~\ref{fig:1dim7}, 
is calculated to be 
\begin{equation}\label{eq:mNu1loop}
 (m_{\nu})_{\alpha\beta} = 
  \frac{N_{c} \sin2\theta_{\rm LQ}}{2(16 \pi^2)}
  \sum_k m_{d_{k}} 
  \Delta B_0(m_{d_{k}}^2,m_1^2,m_2^2) 
  \Big\{
  (\lambda_S)_{\alpha k} (\lambda_D)_{k \beta} 
  + (\alpha\leftrightarrow \beta) \Big\},
\end{equation}
where $N_{c}=3$ is the colour factor.
The loop-integral function $\Delta B_{0}$ 
is given as 
\begin{equation}\label{eq:DelB}
\Delta B_0(m_{d_{k}}^2,m_1^2,m_2^2) 
= \frac{m_1^2 \ln(m_1^2/m_{d_{k}}^2)}{m_1^2-m_{d_{k}}^2} 
- \frac{m_2^2 \ln(m_2^2/m_{d_{k}}^2)}{m_2^2-m_{d_{k}}^2} .
\end{equation}
with 
the eigenvalues $m_{1,2}^2$ of the leptoquark mass matrix 
Eq.~(\ref{eq:massLQ}) and 
the mass $m_{d_{k}}$ of the down-type quark of the $k$-th generation.
Due to the hierarchy in the down-type quark masses, it is expected 
that the contribution from $m_b$ dominates 
the neutrino mass Eq.~(\ref{eq:mNu1loop}). 
For $m_b\ll {\bar M}$ and $\mu v \ll  {\bar M}^2$ 
where ${\bar M} = m_D^2 = m_S^2$, 
Eq.~(\ref{eq:mNu1loop}) is reduced to
\begin{equation}\label{eq:massLQ2}
(m_{\nu})_{\alpha\beta} = \frac{3}{16 \pi^2}\frac{\mu v}{{\bar M}^2}
 m_{b}
 \Big\{
 (\lambda_S)_{\alpha 3} (\lambda_D)_{3 \beta} 
 + (\alpha \leftrightarrow \beta) 
 \Big\},
\end{equation}
and this gives roughly
\begin{equation}\label{eq:massLQ3}
(\lambda_S)_{e 3} (\lambda_D)_{3 e}
=
10^{-2}
\left(
\frac{\meff}{0.2 {\rm eV}} 
\right) 
\left(
\frac{\rm 1 MeV}{\mu}
\right)
\left(
 \frac{{\bar M}}{\rm 1 TeV}
\right)^2.
\end{equation}
The constraint on the effective neutrino mass $\meff \lesssim 0.2$ eV
is derived from the combined KamLAND-Zen and EXO
data~\cite{Gando:2012zm}, which is $T_{1/2} \ge 3.4 \times 10^{25}$ ys
for $^{136}$Xe.  The same experimental results also constrain the
coefficient of the $d=7$ operator generated from the Lagrangians
Eqs.~\eqref{eq:lagLQ} and \eqref{eq:lagLQH} as $\epsilon_{T_R}^{T_R}
\lsim 5.6 \times 10^{-10}$ (cf. Table \ref{Tab:LimitsLong}), which
gives
\begin{align}\label{eq:eps3}
(\lambda_S)_{e 1} (\lambda_D)_{1 e}
\lesssim&
3 \cdot 10^{-2}
\left(
\frac{1{\rm MeV}}{\mu}
\right)
\left(
\frac{\bar{M}}{1{\rm TeV}}
\right)^{4}
\end{align}
Therefore, for $(\lambda_S)_{e 1} (\lambda_D)_{1 e} \simeq 
(\lambda_S)_{e 3} (\lambda_D)_{3 e}$, 
the mass mechanism and the $d=7$ contribution are
approximately of equal size with ${\bar M} \simeq 750$ GeV. 
Since $\meff \propto{\bar M}^{-2}$, while $\epsilon_{{\cal O}_{3,\#2}}
\propto {\bar M}^{-4}$, 
the mass mechanism will dominate $\znbb$ decay 
for ${\bar M}$ larger than ${\bar M}\simeq 750$ GeV, 
unless the couplings $(\lambda_S)_{e 1} (\lambda_D)_{1 e}$ 
are larger than $(\lambda_S)_{e 3} (\lambda_D)_{3 e}$.
We note that, leptoquark searches by the ATLAS
\cite{Stupak:2012aj,ATLAS:2012aq} and the CMS
\cite{CMS:2014qpa,Khachatryan:2015bsa,Khachatryan:2014ura}
collaborations have provided lower limits on the masses of the scalar
leptoquarks, depending on the lepton generation they couple to and
also on the decay branching ratios of the leptoquarks.  The limits
derived from the search for the pair-production of leptoquarks are
roughly in the range $650-1000$ GeV
\cite{Stupak:2012aj,ATLAS:2012aq,CMS:2014qpa,Khachatryan:2015bsa,
  Khachatryan:2014ura}, depending on assumptions.

The other 1-loop models are qualitatively similar to the example
discussed above. However, the numerical values for masses and
couplings in the high-energy completions should be different,
depending on the Lorentz structure of the $d=7$ operators, see also
the appendix.

\subsection{Two-loop level}
\label{subsect:2lp}

We now turn to a discussion of 2-loop neutrino mass models.  
As shown in the appendix, 
in case of the operators ${\cal O}_3$ and ${\cal O}_4$, 
2-loop models appear only for the cases ${\cal O}_{3a}$ 
and ${\cal O}_{4b}$. 
As explained in section \ref{sect:setup}, these operators alone 
cannot give realistic neutrino mass models. 
We thus base our example model on ${\cal O}_8$.
The 2-loop neutrino mass models based on ${\cal O}_{8}$ 
are listed in Tab.~\ref{Tab:BL8} in the appendix. 
In this section, we will discuss decomposition \#15 in detail,
which has not been discussed in the literature before.

\begin{figure}[t]
\begin{center}
\unitlength=1cm
 \begin{picture}(17.5,5.5)
\put(0,0.5){\includegraphics[width=0.35\linewidth]{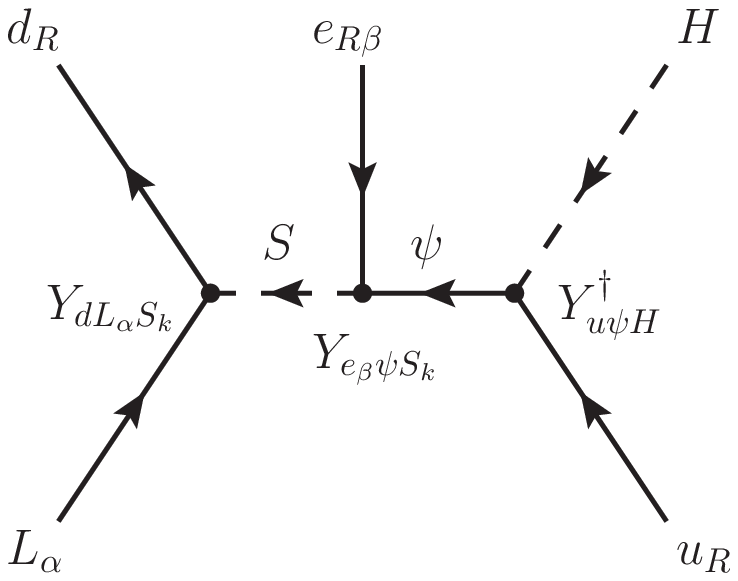}}
\put(6.5,0){\includegraphics[width=0.65\linewidth]{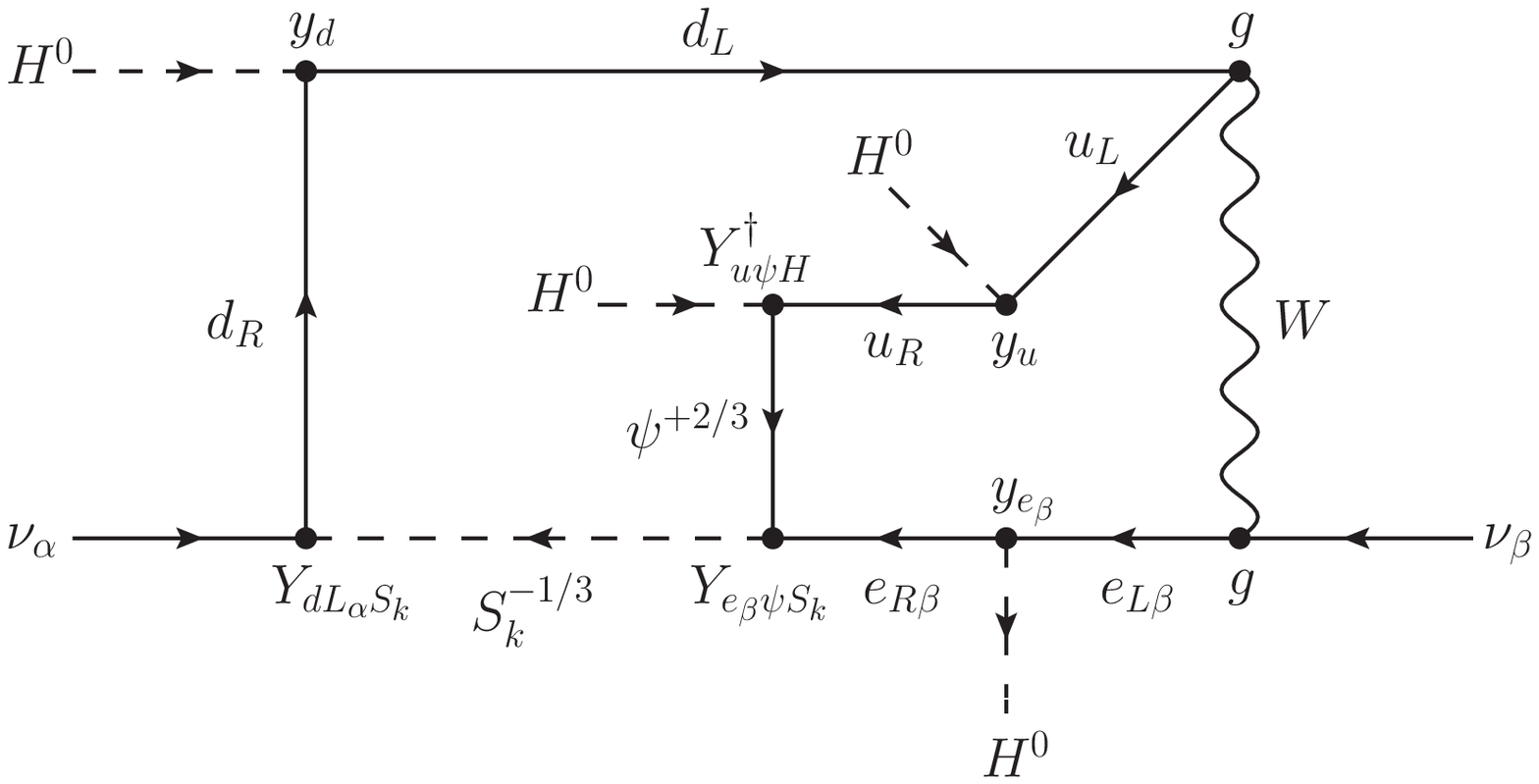}}
 \end{picture}
\caption{Decomposition \#15 of ${\cal O}_{8}$ operator (left)
 and two-loop diagram for neutrino masses based on the 
 decomposition (right).}
\label{fig:2dim7}
\end{center}
\end{figure}

In this model, 
we add the following states to the SM particle content: 
\begin{align}
(\psi_{L,R})_{3,2,7/6}
 =&
 \begin{pmatrix}
  \psi_{L,R}^{5/3}
  \\
  \psi_{L,R}^{2/3}
 \end{pmatrix},
\\
 (S_{3,2,1/6})_k
  =&
 \begin{pmatrix}
  S_k^{2/3}
  \\
  S_k^{-1/3}
 \end{pmatrix}.
\end{align}
With the new fields, we have the interactions 
\begin{align}
 \mathcal{L}_{8,\#15}
 =&
 Y_{d_i L_{\alpha} S_{k}}
 \overline{d_{R,i}} L_{\alpha} \cdot S_{k}
 +
 Y_{u_i \psi H}
 \overline{u_{R,i}} \psi_{L} H^{\dagger}
 +
 Y_{e_{\alpha} \psi S_{k}} \overline{{e_{R \alpha}}^{c}}  \psi_{R}
 S_{k}^{\dagger}
 +
 {\rm h.c.},
\end{align}
which mediate ${\cal O}_{8}$ operator, 
as shown in the left diagram of Fig.~\ref{fig:2dim7}.
Here, $i$ runs over the three quark generations. While 
$Y_{d_i L_{\alpha} S_{k}}$ and  $Y_{u_i \psi H}$ could be different 
for different $i$, for simplicity we will assume the couplings 
to quarks are the same for all $i$ and drop the index $i$ in 
the following. We will comment below, when we discuss the 
numerical results, on how this choice affects phenomenology.
For simplicity, we introduce only one generation of the new fermion
$\psi$, while we allow for more than one copy of the scalar
$S_{3,2,1/6}$. 
Note that, in principle, the model would work also for one copy of 
$S_{3,2,1/6}$ and more than one $\psi$, but as we will see later,
the fit to neutrino data becomes simpler in our setup.

The fermion $\psi^{2/3}$ mixes with the up-type quarks through the
following mass term:
\begin{align} \label{Mix}
{\cal L}_{\text{mass}}
=&
 Y_{u} \langle H^{0} \rangle 
 \overline{u_{L}} u_{R}
 +
 M_{\psi}
 \overline{\psi^{2/3}_{L}}
 \psi_{R}^{2/3}
 +
 Y_{u \psi H}^{\dagger} 
 \langle H^{0} \rangle
 \overline{\psi_{L}^{2/3}}
 u_{R}
 +
{\rm h.c.},
\\  \nonumber 
=&
\begin{pmatrix}
  \overline{t_{L}}
  &
  \overline{\psi_{L}^{2/3}}
 \end{pmatrix}
 \begin{pmatrix}
  m_{t} & 0 
  \\
  \Delta & M_{\psi}
 \end{pmatrix}
 \begin{pmatrix}
  t_{R}
  \\
  \psi_{R}^{2/3}
 \end{pmatrix}
 +
 {\rm h.c.},
\end{align} 
where $\Delta \equiv Y_{u\psi H}^{\dagger} \langle H^{0} \rangle $. 
%
Due to the strong hierarchy in up-type quark masses, 
we have assumed the sub-matrix for the up-type quarks 
in Eq.~(\ref{Mix}) is completely dominated by the contribution 
from top quarks.  
The mass matrix Eq.~(\ref{Mix}) is diagonalized
with the unitary matrices $V_L$ and $V_R$ as
\begin{align}
\label{Diag}
 V_L^{\dagger}
 \begin{pmatrix}
  m_{t} & 0 
  \\
  \Delta & M_{\psi}
 \end{pmatrix}
 V_{R}
 =
 \text{diag}({M}_{\Psi_{i}}),
\end{align}
and the mass eigenstates $\Psi^{2/3}_{i}$ are give as
\begin{align}
\begin{pmatrix}
 t_{L} 
 \\
 \psi^{2/3}_{L}
\end{pmatrix}_{a}
 =
 (V_{L})_{a i}
 \Psi^{2/3}_{L i}
 \quad
 \text{and}
 \quad
 \begin{pmatrix}
 t_{R} 
 \\
 \psi^{2/3}_{R}
\end{pmatrix}_{a}
 =
 (V_{R})_{a i}
 \Psi^{2/3}_{R i},
\end{align}
where the index $a$ for the interaction basis takes 
$a \in \{t, \psi\}$.
The interactions are written in the mass eigenbasis as follows:
\begin{align}\label{Lag1}
 {\cal L}_{W}
 =&
 \frac{g}{\sqrt{2}}
 {(V_L^\dag)_{i t}}
 \overline{\Psi^{2/3}_i}
 \gamma^{\rho}
 P_{L}
 b 
 W^{+}_{\rho}
 +
 {\rm h.c.},
\\
 {\cal L}_{S} =& 
 Y_{d L_{\alpha} S_k} \overline{d_R} L_{\alpha} \epsilon S_k 
 +  
 Y_{e_{\alpha} \psi S_k} 
 (V_R)_{\psi i} \overline{{e_{\alpha}}^c}
 P_{R} \Psi^{2/3}_{i} S_k^{-1/3 \dag} 
 + 
 Y_{e_{\alpha} \psi S_k} 
 \overline{{e_{\alpha}}^c} P_{R} \psi^{5/3} S_k^{2/3 \dag}
 +
 {\rm h.c.}.
\label{Lag}
\end{align} 
The 2-loop neutrino mass diagram generated by this model is shown in
Fig. \ref{fig:2dim7}. 
Using the formulas given in \cite{Sierra:2014rxa}, 
one can express the neutrino mass matrix as
\begin{eqnarray}\label{Mnu}
(m_{\nu})_{\alpha\beta}
 &=& 
 \frac{N_{c} g^{2} m_{b} (V^\dag_L)_{i t} (V_R)_{\psi i}}
 { 2 (16\pi^2)^{2} \ {M}_{\Psi_{i}}}
 \left[
  m_{e_\alpha} Y_{d L_\beta S_k} Y_{e_\alpha \psi  S_k } + 
  m_{e_\beta}  Y_{d L_\alpha S_k} Y_{e_\beta \psi  S_k } 
      \right]
 I(z_{k,i},r_i,t_i) \label{eq:mnu-2loop-example} .
\end{eqnarray}
Here $N_c = 3$ is the colour factor and 
$I(z_{k,i},r_i,t_{i})$ is the loop integral defined as
\begin{eqnarray}
I(z_{k,i},r_i,t_i) = 
\left[4 {\hat I}(z_{k,i},r_i,t_i) - 
\frac{1}{t_i}{\hat I^{(k^2)}}(z_{k,i},r_i,t_i) \right],
\end{eqnarray}
with
\begin{eqnarray}
{\hat I}(z_{k,i},r_i,t_i)=\frac{1}{\pi^4}\int d^4q \int d^4k 
\frac{1}{(q^2 - z_{k,i})(q^2 - r_i)(k^2 - t_i) k^2 ((q+k)^2 - 1)}, \\
{\hat I^{(k^2)}}(z_{k,i},r_i,t_i)=\frac{1}{\pi^4}\int d^4q \int d^4k 
\frac{k^2}{(q^2 - z_{k,i})(q^2 - r_i)(k^2 - t_i) k^2 ((q+k)^2 - 1)},
\label{eq:int2}
\end{eqnarray}
The dimensionless parameters $z_{k,i}, r_i, t_i$ are defined as 
\begin{eqnarray}
z_{k,i}\equiv\frac{m^2_{S_{k}}}{M^2_{\Psi_{i}}},
\quad\quad 
r_i\equiv\frac{m_b^2}{M^2_{\Psi_{i}}},
\quad\quad \mathrm{and}\quad\quad 
t_i \equiv\frac{M^2_{W}}{M^2_{\Psi_{i}}}
\label{eq:def}
\end{eqnarray}
and loop momenta $q$ and $k$ are also defined dimensionless.  Due to
the strong hierarchy in down-type quark masses, we expect that
neutrino mass given in Eq.~(\ref{eq:mnu-2loop-example}) is dominated
by the contribution from bottom quark.
If we assume in Eq.~(\ref{Mnu}) that all Yukawa couplings are of the
same order, then the entries of the neutrino mass matrix will have a
strong hierarchy: $(m_{\nu})_{ee} : (m_{\nu})_{\mu \mu}
:(m_{\nu})_{\tau \tau} = m_e : m_\mu : m_\tau$. Such a flavor
structure is not consistent with neutrino oscillation data.
Therefore, in order to reproduce the observed neutrino masses 
and mixings, 
our Yukawa couplings need to have a certain compensative hierarchy 
in their flavor structure. 

Since the neutrino mass matrix, and thus the Yukawa couplings
contained in the neutrino mass, have a non-trivial flavour pattern,
these Yukawas will be also constrained by charged lepton flavour
violation (LFV) searches.
Here we discuss only $\mu\to e \gamma$ which usually provides the
most stringent constraints in many models.  In order to calculate the
process $\mu\rightarrow e \gamma$ we adapt the general formulas shown
in \cite{Lavoura:2003xp} for our particular case.  The amplitude for
$\mu\rightarrow e \gamma$ decay is given by
\begin{eqnarray}
{\cal{M}}(\mu \rightarrow e \gamma)=
e 
\epsilon^*_{\alpha}
q_{\beta} 
\bar{u}(p_e) i \sigma^{\alpha\beta}
(\sigma_R P_R + \sigma_L P_L)
u(p_{\mu}).
\end{eqnarray}
Here, $\epsilon_{\alpha}$ is the photon polarization vector and
$q_{\beta}$ is the momentum of photon.
Three different diagrams contribute to the amplitude for
$\mu\rightarrow e \gamma$, which are finally summarized 
with the two coefficients $\sigma_R$ and $\sigma_L$ 
given by
\begin{eqnarray}
\sigma_R=
i 
\frac{m_{\mu}}{16\pi^{2}}
\left[
Y_{d L_2 S_k}^\dagger Y_{d L_1 S_k}
\frac{ 2 F_{2} (x_{b,k}) -  F_{1}(x_{b,k})}{m_{S_{k}}^{2}}
\right],
\label{eq:sigmaigmaR}
\end{eqnarray}
\begin{eqnarray}
\sigma_L=
i 
\frac{m_{\mu}}{16\pi^{2}}
\left[
Y^\dagger_{e_2 \psi S_k} Y_{e_1 \psi S_k}
\frac{-F_{2} (x_{\psi,k}) -7 F_{1}(x_{\psi,k})}{m_{S_{k}}^{2}}
\right],
\label{eq:sigmaigmaL}
\end{eqnarray}
where $x_{\psi,k} \equiv \frac{M^2_{\psi}}{m^2_{S_{k}}}$ and
$x_{b,k}\equiv\frac{m^2_{b}}{m^2_{S_{k}}}$.  
Here, we have 
assumed that both the $\psi^{-2/3}$ and the $\psi^{-5/3}$ 
have the same mass $M_{\psi}$. This neglects (small) mass 
shifts in the $\psi^{-2/3}$ state, due to its mixing with 
the top quark. Due to the large value of $M_{\psi}$, that 
we use in our numerical examples, this should be a good 
approximation. Note also, that the contribution from the 
top quark is negligible for those large values of 
$M_{\psi}$ used below. 
The functions $F_{1}(x)$ and $F_{2}(x)$ are defined in 
Eqs~(40) and (41) in \cite{Lavoura:2003xp} as
\begin{align}
F_{1}(x)=&
 \frac{x^{2} - 5 x -2}{12(x-1)^{3}} + \frac{ x \ln x}{2 (x-1)^{4}},
 \\
F_{2}(x)=&
 \frac{2 x^{2} + 5 x -1}{12 (x-1)^{3}}
 -
 \frac{x^{2} \ln x}{2 (x-1)^{4}},
\end{align}
The branching ratio for $\mu\rightarrow e\gamma$  
can be expressed with the coefficients $\sigma_{R}$ and $\sigma_{L}$ as
\begin{eqnarray}\label{eq:brmueg}
\mathrm{Br}(\mu\rightarrow e \gamma)
=
\frac{e^{2} m_{\mu}^{3} 
(\left| \sigma_{R} \right|^{2} + \left| \sigma_{L} \right|^{2})}{16 \pi \ \Gamma_\mu},
\end{eqnarray}
where $\Gamma_\mu$ is the total decay width of muon.
Later, we will numerically calculate the branching ratio 
to search for the parameter choices that are consistent 
with the oscillation data and the constraint from 
$\mu \rightarrow e \gamma$.

Before discussing constraints from lepton flavour violation, we will
compare the long-range contribution to $\znbb$ with the mass mechanism
in this model.
This model manifestly generates a $d=7$ long-range contribution
to $\znbb$. 
The half-life of $\znbb$ induced by the long-range contribution 
is proportional to the coefficient $\epsilon_{V+A}^{V+A}$
which is expressed in terms of the model parameters as
\begin{eqnarray}\label{eq:eps32}
 \epsilon_{V+A}^{V+A} = 
\frac{\sqrt{2}}{G_F } \frac{\Delta}{M_\psi} \frac{ Y_{e_1 \psi S_k}  Y_{d L_1 S_k}}
{2 m_{S_k}^2}  \lsim 3.9 \times 10^{-7}.
\end{eqnarray}
Here, we use the limit on $ \epsilon_{V+A}^{V+A}$ from non-observation
of $^{136}$Xe $\znbb$ decay, see Table~\ref{Tab:LimitsLong}.  With one
copy of the new scalar, the bound of Eq.~(\ref{eq:eps32}) is directly
related to the effective neutrino mass
Eq.~(\ref{eq:mnu-2loop-example}) and places the stringent constraint:
\begin{eqnarray}\label{eq:limLR}
\meff 
\lesssim
2 \times 10^{-5} \text{[eV]} 
\left(
\frac{m_S}{10 \text{[TeV]}}
\right)^2 
\left(
\frac{ I(z_{k,1},r_{1},t_{1}) }{5 \times 10^{-2}}
\right),
\end{eqnarray}
where we have used the approximate relation 
\begin{eqnarray}\label{eq:app}
 \frac{(V^\dag_L)_{it} (V_R)_{\psi i} I(z_{k,i},r_i,t_i)}
{{M}_{\Psi_{i}}} 
\simeq
\frac{ I(z_{k,1}, r_1, t_1) \Delta}{M_\psi m_t} ,
\end{eqnarray}
with $z_{k,1} = (m_{S_k}/m_t)^2$, $r_{1} = (m_{b}/m_{t})^{2}$, $t_1 =
(M_W/m_t)^2$, and $ I(z_{k,1},r_{1},t_{1}) \sim 5 \times 10^{-2}$ for a
  scalar mass of $m_S = 10 $ TeV and $M_{\psi} \simeq 0.8 $ TeV. 
Note that this parameter choice is motivated by the fact that the
model cannot fit neutrino data with perturbative Yukawa couplings with
scalar masses larger than $m_S \gsim 10$ TeV.  As one can see from
Eq.~(\ref{eq:limLR}), the long-range contribution to $\znbb$ clearly
dominates over the mass mechanism in this setup.

In short, this neutrino mass model predicts large decay rate of
$\znbb$ but tiny $\langle m_{\nu} \rangle$. This implies that, if
future neutrino oscillation experiments determine that the neutrino
mass pattern has normal hierarchy but $\znbb$ is discovered in the
next round of experiments, the $\znbb$ decay rate is dominated 
by the long-range part of the amplitude. Recall that 
${\cal O}_8$ contains $\overline{e^{c}}$. This implies that the 
model predicts a different angular distribution than the 
mass mechanism, which in principle could be tested  
in an experiment such as Super-NEMO \cite{Arnold:2010tu}. 

Note that, to satisfy the condition Eq.~\eqref{eq:limLR},
cancellations among different contributions to $\meff$ are necessary.
This can be arranged only if we consider at least two generations of
the new particles in the model (either the scalar $S$ or the fermion
$\psi$).

Here we discuss more on the consistency of our model with the neutrino
masses and mixings observed at the oscillation experiments.  Instead
of scanning whole the parameter space, we illustrate the parameter
choice that reproduces the neutrino properties and is simultaneously
consistent with the bound from lepton flavour violation.  To simplify
the discussion we use the following ansatz in the flavour structure of
the Yukawa couplings:
\begin{equation}\label{anz} 
Y_{d L_\alpha S_k}
=
\frac{Y_{e_\alpha \psi S_k}}{y}
\frac{m_{e_\alpha}}{m_\mu} 
\end{equation}
with a dimensionless parameter $y$.
With Eq.~\eqref{anz},
the neutrino mass matrix Eq.~\eqref{eq:mnu-2loop-example}
is reduced to 
\begin{eqnarray}
(m_{\nu})_{\alpha\beta} 
=  
(\Lambda)_{\alpha k}
\mathcal{I}_{k}
(\Lambda^{T})_{k \beta},
\label{eq:mnuCI}
\end{eqnarray}
where $\Lambda$ is defined as
\begin{eqnarray}
\label{lambda}
\Lambda_{\alpha k} 
\equiv 
Y_{d L_\alpha S_k}
=
\frac{Y_{e_\alpha \psi S_k}}{y}\frac{m_{e_\alpha}}{m_\mu},  
\end{eqnarray}
and $\mathcal{I}$ is given as
\begin{eqnarray}
\mathcal{I}_{k}
=
\frac{N_{c} g^2 y m_{b} m_{\mu} (V^\dag_L)_{i t} (V_R)_{\psi i}}
{ (16\pi^2)^{2} M_{\Psi_{i}} } I(z_{k,i},r_i,t_i).
\end{eqnarray}
We introduce three copies of the new scalar $S^{-1/3}_k$.
The resulting mass matrix Eq.~\eqref{eq:mnuCI} has the same index
structure as that of the type-I seesaw mechanism, and therefore, the
matrix $\Lambda$ can be expressed as
\begin{eqnarray}
\label{eq:CI}
\left( \Lambda^{T} \right)_{k \alpha}
=
\left( \sqrt{\mathcal{I}^{-1}} \right)_{k} 
R_{ki} 
\left( \sqrt{\hat{m}_{\nu}} \right)_{i}
\left( U_{\nu}^{\dagger} \right)_{i \alpha},
\end{eqnarray}
following the parameterization developed 
by Casas and Ibarra~\cite{Casas:2001sr}.

Here, $\hat{m}_{\nu}$ is the neutrino mass matrix in the mass
eigenbasis, and the mass matrix $m_{\nu}$ 
is diagonalized with the lepton mixing matrix 
$U_{\nu}$ as
\begin{equation}
(\hat{m}_{\nu})_{i}
\equiv 
\text{diag}
\begin{pmatrix}
 m_{\nu_{1}}
 &
 m_{\nu_{2}}
 &
 m_{\nu_{3}}
\end{pmatrix}
=
(U_{\nu}^{T})_{i \alpha} \, 
(m_{\nu})_{\alpha \beta} \, 
(U_{\nu})_{\beta j} 
\end{equation}
for which we use the following standard parametrization
\begin{equation}
\label{eq:mixing}
U_{\nu}=
\left(
\begin{array}{ccc}
 c_{12}c_{13} & s_{12}c_{13}  & s_{13}e^{i\delta}  \\
-s_{12}c_{23}-c_{12}s_{23}s_{13}e^{-i\delta}  & 
c_{12}c_{23}-s_{12}s_{23}s_{13}e^{-i\delta}  & s_{23}c_{13}  \\
s_{12}s_{23}-c_{12}c_{23}s_{13}e^{-i\delta}  & 
-c_{12}s_{23}-s_{12}c_{23}s_{13}e^{-i\delta}  & c_{23}c_{13}  
\end{array}
\right) 
\left(
\begin{array}{ccc}
1 & 0 & 0 \\
0 & e^{i\alpha_{21}}  & 0 \\
0 & 0 & e^{i\alpha_{31}}
\end{array}
\right).
\end{equation}
Here $c_{ij} =\cos \theta_{ij}$, $s_{ij} = \sin \theta_{ij}$ with the
mixing angles $\theta_{ij}$, $\delta$ is the Dirac phase and
$\alpha_{21}$, $\alpha_{31}$ are Majorana phases.  
The matrix $R$ is a
complex orthogonal matrix which 
can be parametrized in terms of three complex angles as
\begin{equation} \label{RR}
R\ =\ \left( \begin{array}{ccc} 
c_{2} c_{3} & -c_{1} s_{3}-s_1 s_2 c_3& s_{1} s_3- c_1 s_2 c_3\\ 
c_{2} s_{3} & c_{1} c_{3}-s_{1}s_{2}s_{3} & -s_{1}c_{3}-c_1 s_2 s_3 \\ 
s_{2}  & s_{1} c_{2} & c_{1}c_{2}\end{array} \right).
\end{equation}
Note that it is assumed in this procedure 
that the charged lepton mass matrix is diagonal.
After fitting the neutrino oscillation data with the parametrization
shown above, there remain $y$, $Y_{u\psi H}$ and the masses
$M_{\psi}$, $m_{S_{k}}$ for $k= 1,2,3$ as free parameters.  For
simplicity, we assume a degenerate spectrum of the heavy scalars
$m_{S} = m_{S_{k}}$.

\begin{figure}[t]
\centering
\includegraphics[width=0.48\linewidth]{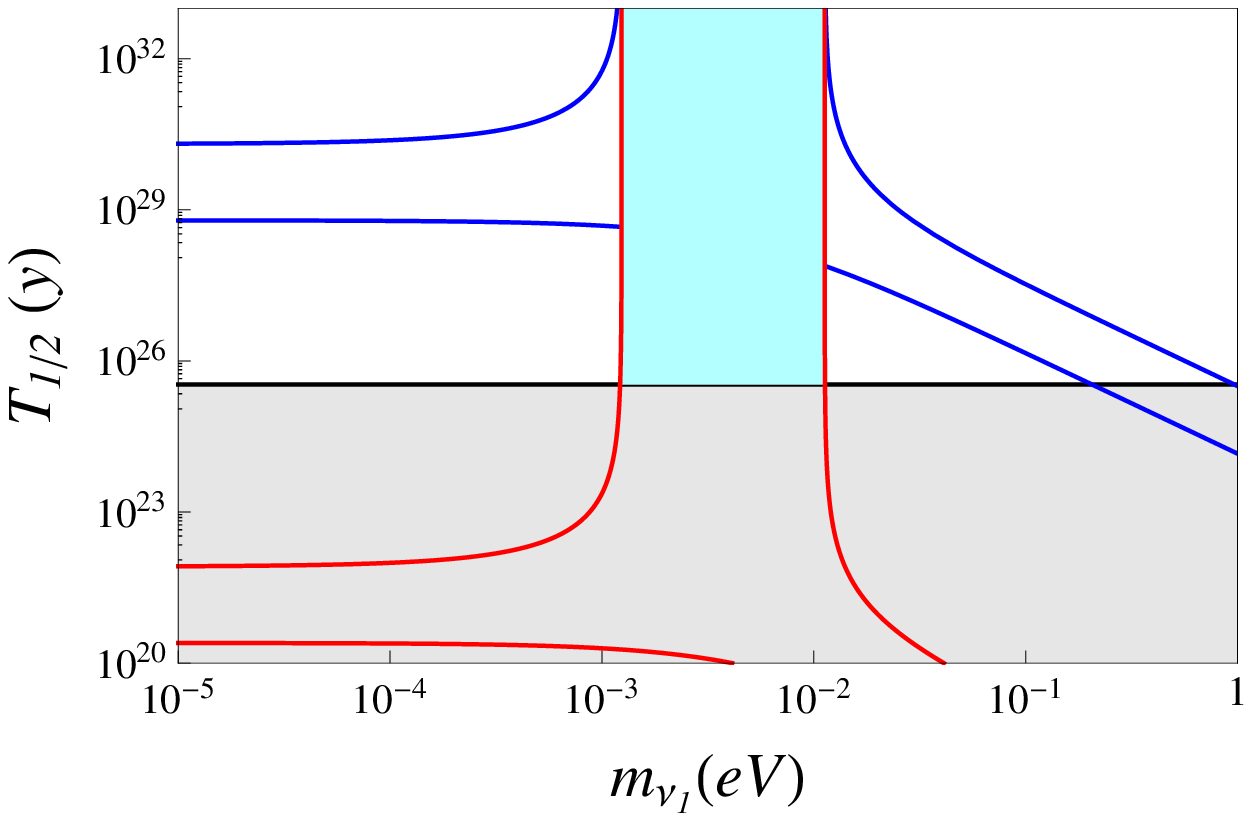}
\includegraphics[width=0.49\linewidth]{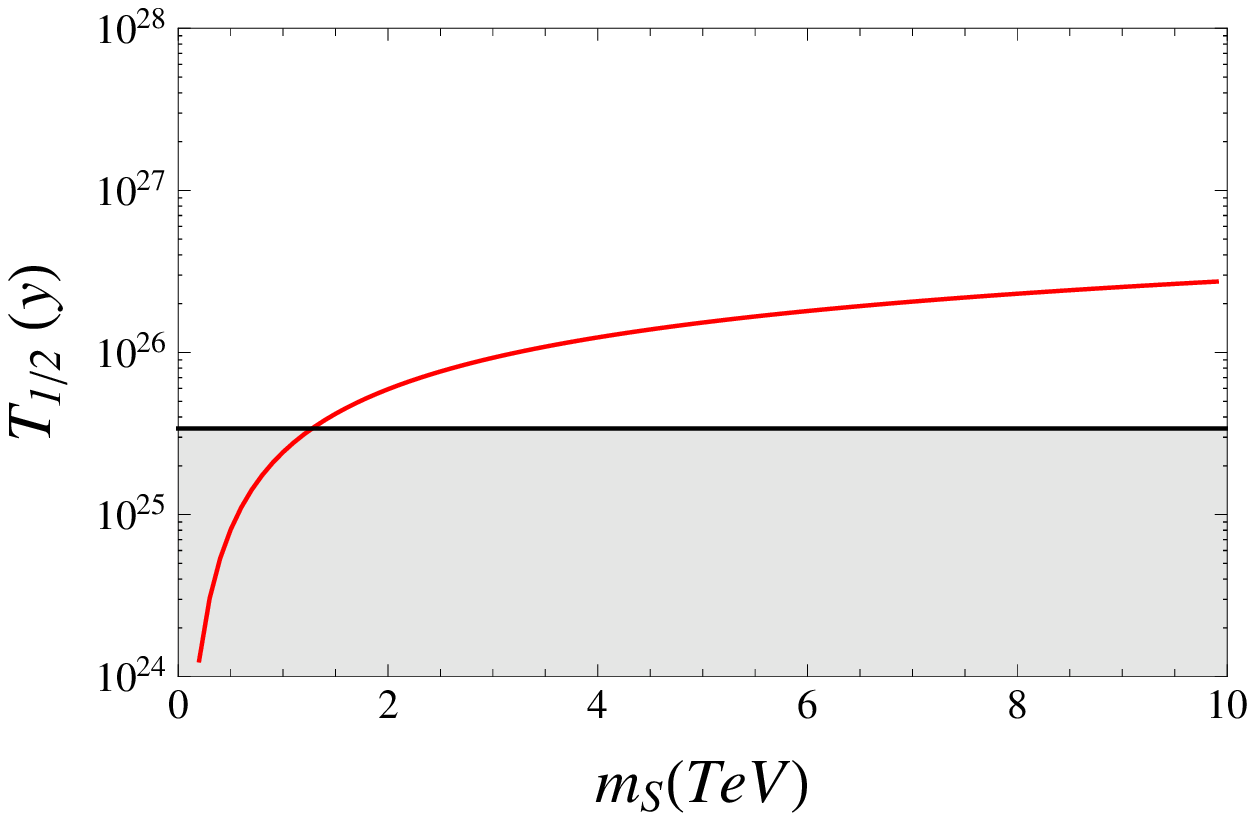}
\caption{Calculated half-lives for $\znbb$ decay of $^{136}$Xe
  considering the long-range contribution to the decay rate versus
  $m_{\nu_1}$ (left) and $m_S$ (right). The gray region is the current
  lower limit in $\znbb$ decay half-life of $^{136}$Xe.  In the plot
  to the left the region between the red curves is the one allowed by
  the long-range contribution to the decay rate of $\znbb$ calculated
  scanning over oscillation parameters for the case of normal
  hierarchy and $m_S = 10 \ \text{TeV}$. We also show the allowed
  region for the half-live for the mass mechanism as blue lines for
  comparison. The cyan region correspond to the parametric region
  where our model can be consistent with current $\znbb$ experimental
  data. In the plot to the right the red curve is the long-range
  contribution to the decay rate for the fixed oscillation parameters
  $m_{\nu_1}=1.23 \times 10^{-3}$ eV , $\alpha_{21} = 0$, $\alpha_{31}
  = \pi/2$, $s_{23}^2=1/2$ and $s_{12}^2=1/3$ and the remaining
  oscillation parameters $\Delta m_{31}^2$ and $\Delta m_{21}^2$ fixed
  at their best-fit values for the case of normal hierarchy.  }
\label{fig:tlr}
\end{figure}
In Fig.~\ref{fig:tlr}-(a), 
we plot the half-life $T_{1/2}^{\znbb}$
as a function of $m_{\nu_1}$ for fixed values 
of the coupling $Y_{u \psi H} = 0.6$ and the masses
$M_{\psi} = 800 \ \text{GeV}$ and $m_{S} = 10$ TeV. 
The parameter $y$ is taken to be $10^{-3}$, since this minimizes the
decay rate of $\mu \rightarrow e \gamma$, as we will discuss below.
We have used oscillation parameters for the case of normal hierarchy.
The region enclosed by the red curves is $d=7$ long-range contribution
to $\znbb$, and the blue curves correspond to the mass mechanism
contribution only, which is shown for comparison.  The gray region is
already excluded by $\znbb$ searches, and for the model under
consideration only the cyan region is allowed.  As one can see from
Fig.~\ref{fig:tlr}-(a), the total contribution to $\znbb$ is dominated
by the $d=7$ long-range contribution.  Note that the mass mechanism
and the long-range contribution are strictly related only under the
assumption that $Y_{u \psi H}$ and $Y_{d L_{\alpha} S_{k}}$ are
independent of the quark generation $i$. This is so, because the
2-loop diagram is dominated by 3rd generation quarks, while in $\znbb$
decay only first generation quarks participate. If we were to drop
this assumption and put the first generation couplings to $Y_{u_1 \psi
  H} \lsim 10^{-2} \times Y_{u_3 \psi H}$ and $Y_{d_1 L_{\alpha}
  S_{k}} \lsim 10^{-2}\times Y_{d_3 L_{\alpha} S_{k}}$, the half-life
for the long-range amplitude would become comparable to the mass
mechanism, without changing the fit to oscillation data.

Note that non-zero Majorana phases are necessary to allow for
cancellations among the mass mechanism contributions, so as to make
$\meff$ small as required by Eq.~(\ref{eq:limLR}). 
In Fig.~\ref{fig:tlr}-(b), we plot the half-life $T_{1/2}^{\znbb}$
as a function of the scalar mass $m_S$. 
Here we fixed the oscillation parameters to $m_{\nu_1}=1.23 \times
10^{-3}$ eV , $\alpha_{21} = 0$, $\alpha_{31} = \pi/2$, $s_{23}^2=1/2$
and $s_{12}^2=1/3$ and the remaining oscillation parameters $\Delta
m_{31}^2$ and $\Delta m_{21}^2$ to their best-fit values for the case
of normal hierarchy.  The plot assumes that the matrix $R$ is equal to
the identity.  The plot shows that the half-life increases to reach
approximately $T_{1/2}^{\znbb} \sim 10^{26}$ yr for $m_S = 10$ TeV.

\begin{figure}[t]
\centering
\includegraphics[width=0.49\linewidth]{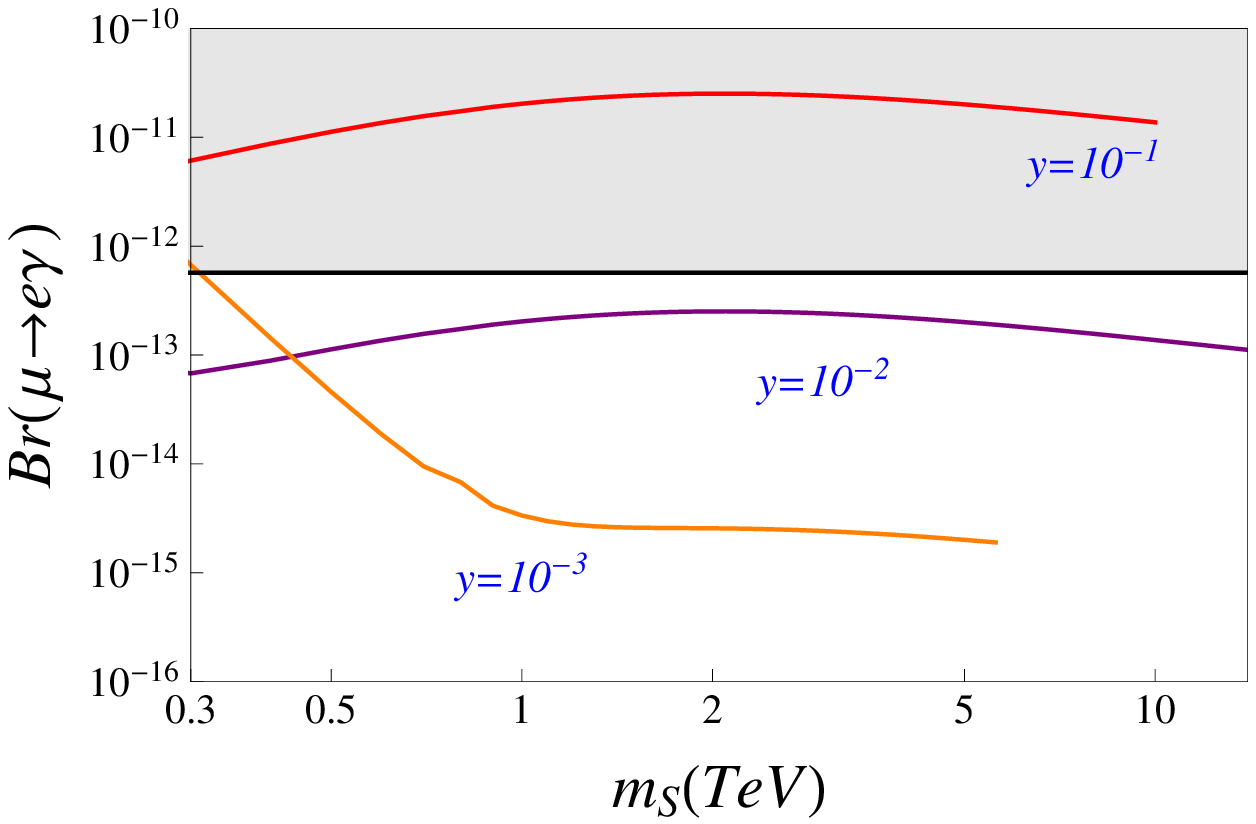}
\includegraphics[width=0.49\linewidth]{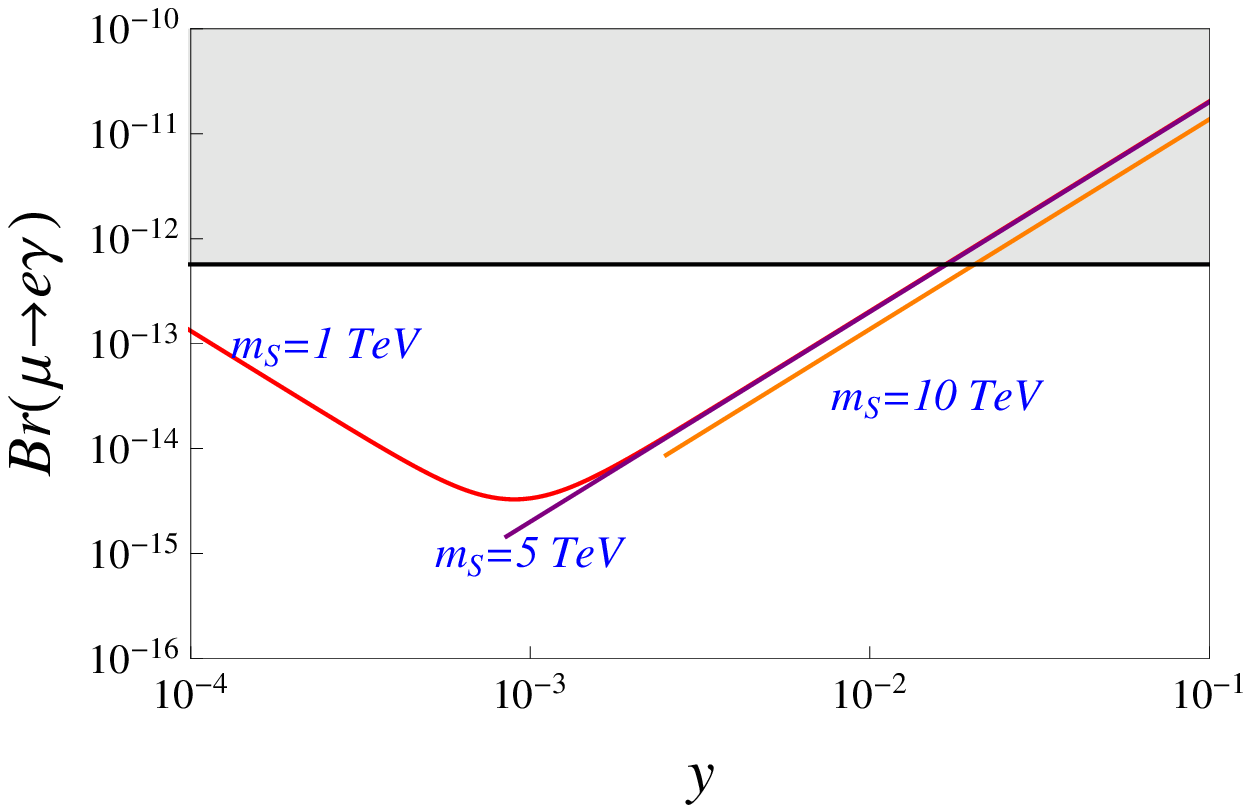}
\caption{Br($\mu\to e\gamma$) versus the scalar $m_S$ (left) and the
  parameter $y$ (right).  In the plot to the left the red, orange and
  purple solid curves are the Br($\mu\to e\gamma$) for different
  values of $y = 10^{-1}, 10^{-2}, 10^{-3}$.  The gray region is the
  current experimental upper limit on Br($\mu\to e\gamma$) from the
  MEG experiment~\cite{Adam:2013mnn}. In the plot to the right the
  red, orange and purple solid curves are the Br($\mu\to e\gamma$) for
  different values of the mass $m_S = 1, 5, 10$ TeV.  We have fixed
  the oscillation parameters to $m_{\nu_1}=1.23 \times 10^{-3}$ eV ,
  $\alpha_{21} = 0$, $\alpha_{31} = \pi/2$, $s_{23}^2=1/2$ and
  $s_{12}^2=1/3$. The remaining oscillation parameters $\Delta
  m_{31}^2$ and $\Delta m_{21}^2$ are fixed at their best-fit values
  for the case of normal hierarchy. For discussion see text. }
\label{fig:br}
\end{figure}
Now we discuss the constraint from lepton flavour violating process 
$\mu\to e\gamma$.  
In Fig.~\ref{fig:br},
we show Br($\mu\to e\gamma$) as a function of the scalar $m_{S}$ and
the parameter $y$ for fixed values of the coupling $Y_{u\psi H} = 0.6$
and the fermion mass $M_{\psi} = 800 \ \text{GeV}$,
which is the same parameter choice adopted in Fig.~\ref{fig:tlr}. 
These plots show that the current experimental limits on 
Br($\mu\to e\gamma$) put strong constraints on the model 
under consideration.
In Fig.~\ref{fig:br}-(a), we plot Br$(\mu \rightarrow e \gamma)$ 
with different values of the parameter 
$y = \{ 10^{-1}, 10^{-2}, 10^{-3} \}$.
We have used again the parameters $m_{\nu_1}=1.23 \times 10^{-3}$
eV, $\alpha_{21} = 0$, $\alpha_{31} = \pi/2$, $s_{23}^2=1/2$ and
$s_{12}^2=1/3$ fixing the remaining oscillation parameters 
$\Delta m_{31}^2$ and $\Delta m_{21}^2$ at their best-fit values 
for the case of normal hierarchy.  
With the choice of $y=10^{-1}$, the entire region of $m_S$ is not
consistent with the current experimental limits.  On the other hand,
we can easily avoid the constraint from $\mu \rightarrow e \gamma$ by
setting the parameter $y$ to be roughly smaller than $10^{-2}$.
Note that the curves with $y=10^{-1}$ and $y=10^{-3}$ 
do not cover the full range of $m_{S}$. 
This is because 
the fit to neutrino data would require Yukawa couplings 
in the perturbative regime. (We define the boundary 
to perturbativity as at least one entry in the Yukawa
matrix being smaller than $\sqrt{4 \pi}$.)
It is necessary to have smaller values of the parameter $y$ to obey
the experimental bound. This feature is also shown in
Fig.~\ref{fig:br}-(b) where we plot the Br($\mu\to e\gamma$) as a
function of $y$ with different values of the mass $m_S = \{1, 5, 10
\}$ TeV.
As shown, for $y \lesssim 10^{-2}$ it is possible to fulfil the
experimental limit, having the Br($\mu\to e\gamma$) a minimum around
$y = 10^{-3}$.
Because of the perturvative condition, the curves with $m_{S}=5$ TeV
and $m_{S}=10$ TeV end in the middle of the $y$ space.  The reason for
the strong dependence of Br($\mu\to e\gamma$) on the parameter $y$ can
be understood as follows: As shown in Eq.~(\ref{lambda}) the Yukawa
couplings $Y_{d L_\alpha S_k}$ and $Y_{e_\alpha \psi S_k}$ are related
in the neutrino mass fit, but only up to an overall constant,
$\frac{1}{y}$. For values of $y$ of the order of $10^{-3}$ both
Yukawas are of the same order and this minimizes Br($\mu\to
e\gamma$). If $y$ is much larger (much smaller) than this value $Y_{d
  L_\alpha S_k}$ ($Y_{e_\alpha \psi S_k}$) becomes much larger than
$Y_{e_\alpha \psi S_k}$ ($Y_{d L_\alpha S_k}$) and since the different
diagrams contributing to Br($\mu\to e\gamma$) are proportional to the
individual Yukawas (and not their product) this leads to a much larger
rate for Br($\mu\to e\gamma$).

In summary, for all 2-loop $d=7$ models of neutrino mass, which lead
to ${\cal O}_8$, the long-range part of the amplitude will dominate
over the mass mechanism by a large factor, unless there 
is a strong hierarchy between the non-SM Yukawa couplings to the 
first and third generation quarks. 
Such models are severely constrained by lepton flavour violation 
and $\znbb$ decay. We note again, that these models predict an 
angular correlation among the out-going electrons which is 
different from the mass mechanism.

\section{Left-right symmetric model: $d=7$ versus $d=9$ operator}
\label{sect:lim}

Writing new physics contributions to the SM in a series of NROs
assumes implicitly that higher order operators are
suppressed with respect to lower order ones by additional inverse
powers of the new physics scale $\Lambda$.  However, there are some
particular example decompositions for (formally) higher-order
operators, where this naive power counting fails.  We will discuss
again one particular example in more detail. The example we choose 
describes the situation encountered in left-right symmetric extensions 
of the standard model.

\begin{figure}[t]
\hskip-2mm\includegraphics[width=0.5\linewidth]{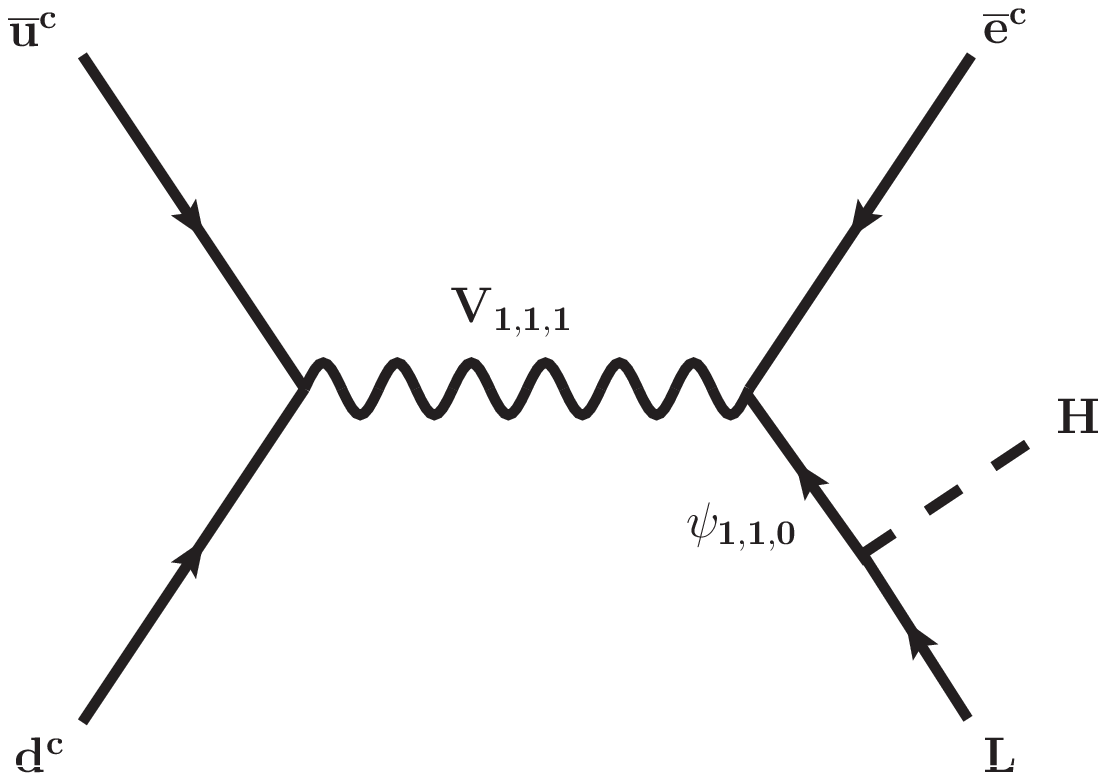}
\includegraphics[width=0.5\linewidth]{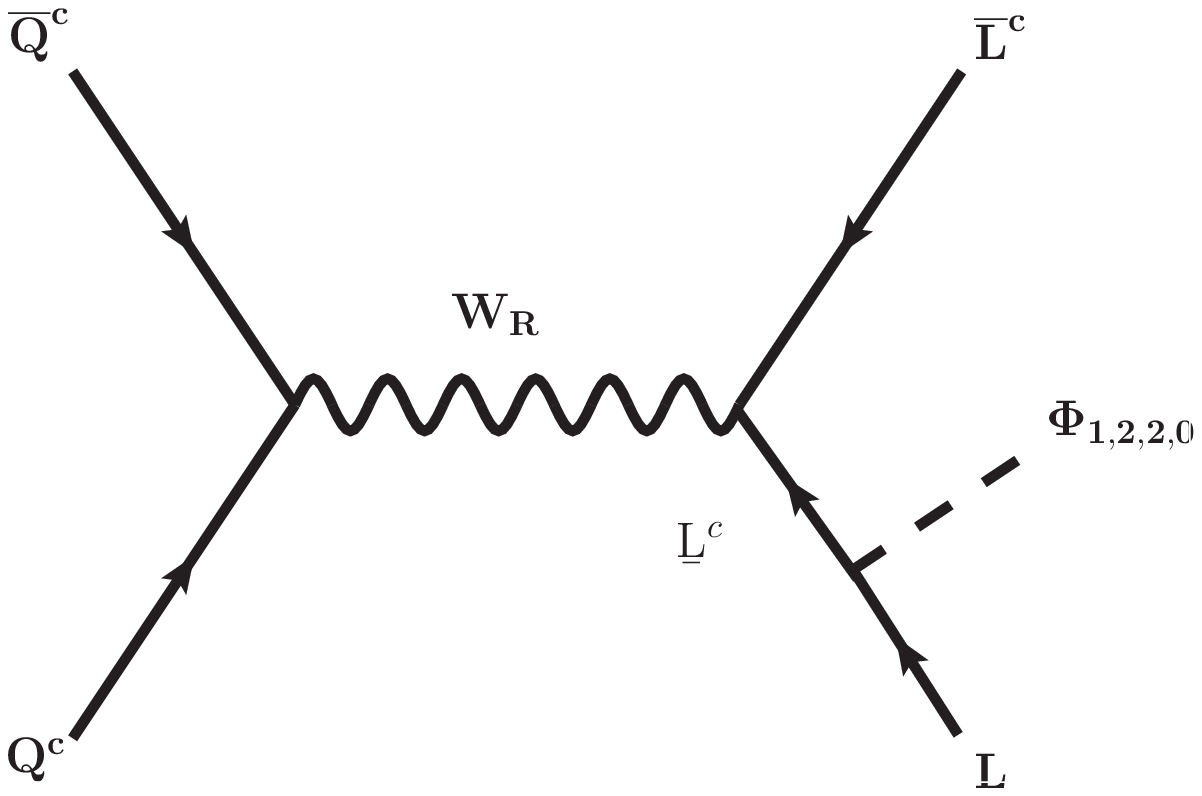}
\caption{${\cal O}_8$ decomposed as \#14: $({\overline u^c}d^c)
({\overline e^c})(LH)$ under the SM gauge group 
(left) and for the LR gauge group (right).}
\label{fig:O8}
\end{figure}

Consider the following two Babu-Leung operators:
\begin{eqnarray}
{\cal O}_8 = L^i \overline{e^c} \hspace{0.05cm} \overline{u^c} 
d^c H^j \epsilon_{ij} 
& & 
{\cal O}_7 = L^i Q^j \overline{e^c} \overline{Q}_k 
H^k H^l H^m \epsilon_{il} \epsilon_{jm} 
\end{eqnarray}
${\cal O}_8$ can be decomposed in a variety of ways, decomposition
\#14 (see Table \ref{Tab:BL8}) is shown in Fig.~\ref{fig:O8} to the
left. 
The charged vector appearing in this diagram couples to a pair of
right-handed quarks and, thus, can be interpreted as the charged
component of the adjoint of the left-right symmetric (LR) extension of
the SM, based on the gauge group $SU(3)_C\times SU(2)_L \times SU(2)_R
\times U(1)_{B-L}$.  In LR right-handed quarks are doublets, $Q^c
=\Psi_{\bar{3},1,2,-1/6}$, the $\psi_{1,1,0}$ can be understood as the
neutral member of $L^c$, i.e. the right-handed neutrino, and the Higgs
doublet is put into the bidoublet, $\Phi_{1,2,2,0}$. The resulting
diagram for $\znbb$ decay is shown in Fig.~\ref{fig:O8} on the
right.

Fig.~\ref{fig:O8} gives a long-range contribution to $\znbb$ decay.
We can estimate the size of $\epsilon_{{\cal O}_8}$ from these
diagrams:
\begin{eqnarray}\label{eq:epsO8}
\frac{G_F\epsilon_{{\cal O}_{8,\# 14}}}{\sqrt{2}} = 
\frac{Y_{L\psi}g_1g_2v_{\rm SM}}{m_V^2m_{\psi}} =
\frac{Y_{LL^c}g_R^2v_u}{m_{W_R}^2m_{\nu_R}} 
\end{eqnarray}
The first of these two equations shows $\epsilon_{{\cal O}_8}$ for
Fig.~\ref{fig:O8} on the left (notation for SM gauge
  group), the second for Fig.~\ref{fig:O8} on the right (notation for
gauge group of the LR model).  Here, $g_1$ and $g_2$ could be
different, in principle, but are equal to $g_R$ in the LR model.
$v_{\rm SM}$ is the SM vev, fixed by the $W$-mass. In the LR model,
the bi-doublet(s) contain in general two vevs. We call them $v_d$ and
$v_u$ here and $v_{\rm SM}^2 = v_d^2 +v_u^2$.  In Eq.~(\ref{eq:epsO8})
only $v_u = v_{\rm SM}\sin\beta$, with $\tan\beta=v_u/v_d$, appears. Note
that we have suppressed again generation indices and summations in
Eq.~(\ref{eq:epsO8}). We will come back to this important point below.

\begin{figure}[t]
\hskip-2mm\includegraphics[width=0.5\linewidth]{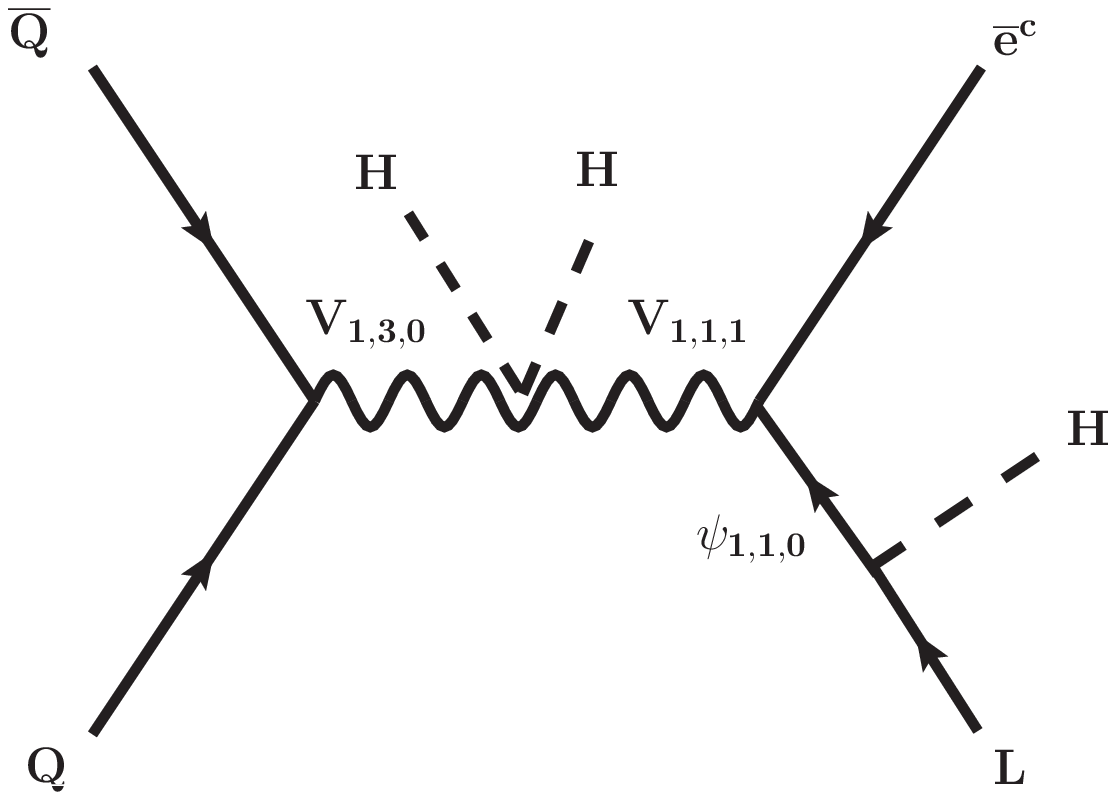}
\includegraphics[width=0.5\linewidth]{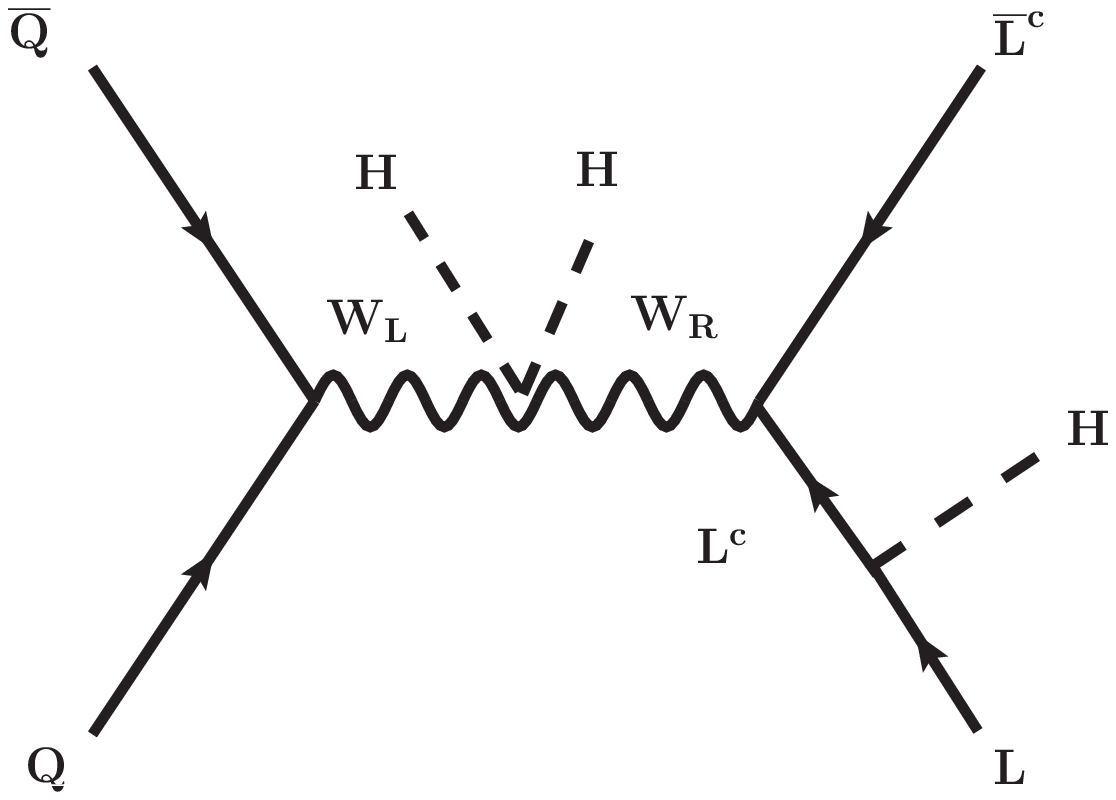}
\caption{${\cal O}_7$ $d=9$ contribution to $\znbb$ decay decomposed as 
$({\overline Q}Q)(HH)({\overline e^c})(LH)$ in the SM (left) and 
in the LR model (right).}
\label{fig:d9}
\end{figure}

Now, however, first consider ${\cal O}_7$. From the many different 
possible decompositions we concentrate on the one shown in 
Fig.~\ref{fig:d9}. The diagram on the left shows the diagram in SM 
notation, the diagram on the right is the corresponding LR 
embedding. It is straightforward to estimate the size of these 
diagrams as:
\begin{eqnarray}\label{eq:epsO7}
\frac{G_F\epsilon_{{\cal O}_7}}{\sqrt{2}} = 
\frac{Y_{L\psi}g_1g_2g_3^2v_{\rm SM}^3}{m_{V_{1,3,0}}^2m_{V_{1,1,1}}^2m_{\psi}} =
\frac{Y_{LL^c}g_L^2g_R^2v_u^2v_d}{m_{W_R}^2m_{W_L}^2m_{\nu_R}} 
\propto \frac{Y_{LL^c}g_R^2v_u}{m_{W_R}^2m_{\nu_R}} 
\end{eqnarray}
Arbitrarily we have called the 4-point coupling in the left diagram
$g_3^2$. 
In the LR model again the couplings are fixed to $g_L$ and
$g_R$. 
In the last relation in Eq.~(\ref{eq:epsO7}) we have used
$v_{\rm SM}^2 \propto m_{W_L}^2/g_L^2$. 
This shows that Eq.~(\ref{eq:epsO7}) is of the same order 
than Eq.~(\ref{eq:epsO8}), {\em despite coming from a $d=9$ operator}.  
This a priori counter-intuitive
result is a simple consequence of the decomposition containing the SM
$W_L$ boson. 
Any higher-order operator which can be decomposed in such
a way will behave similarly, i.e.  $1/\Lambda^5 \Rightarrow
1/(\Lambda^3 v_{\rm SM}^2)$.\footnote{%
In addition to the case of the SM W-boson, 
discussed here, similar arguments apply to decompositions containing 
the scalar $S_{1,2,1/2}$, which can be interpreted as the SM Higgs boson.}

We note that in this particular example the contribution of 
${\cal O}_7$ is actually {\em more stringently} constrained than 
the one from ${\cal O}_8$. 
This is because ${\cal O}_8$ leads to a low-energy
current of the form ($V+A$) in both, the leptonic {\em and} the
hadronic indices, i.e. the limit corresponds to
$\epsilon^{V+A}_{V+A}$.  
${\cal O}_7$, on the other hand, leads to
$\epsilon^{V+A}_{V-A}$, which is much more tightly constraint due to
contribution from the nuclear recoil matrix element \cite{Doi:1985dx}, 
compare values in Table \ref{Tab:LimitsLong}.

We note that, one can identify the diagrams in Fig.~\ref{fig:O8} and
Fig.~\ref{fig:d9} with the terms proportional to $\lambda$ and
$\eta$ in the notation of \cite{Doi:1985dx}, used by many authors in
$\znbb$ decay. 
For the complete expressions for the long-range part of
the amplitude, one then has to sum over the light neutrino mass
eigenstates, taking into account that the leptonic vertices in the
diagrams in Figs.~\ref{fig:O8} and \ref{fig:d9} are right-handed.
Defining the mixing matrices for light and heavy neutrinos as $U_{\alpha j}$ 
and $V_{\alpha j}$, respectively, as in \cite{Doi:1985dx}, the 
coefficients  $\epsilon_{{\cal O}_8}$ and $\epsilon_{{\cal O}_7}$ of 
the $d=7$ and $d=9$ operators are then the {\em effective} 
couplings \cite{Doi:1985dx}:
\begin{eqnarray}\label{eq:effeta}
\langle\lambda\rangle = \sum_{j=1}^3 U_{ej}V_{ej}\lambda & , & 
\langle\eta\rangle = \sum_{j=1}^3 U_{ej}V_{ej}\eta .
\end{eqnarray}
Orthogonality of $U_{ej}$ and $V_{ej}$ leads to $\sum_{j=1}^6
U_{ej}V_{ej} \equiv 0$. 
However, the sum in Eq.~(\ref{eq:effeta}) runs
only over the light states, which does not vanish exactly, 
but rather is expected to be of the order of the light-heavy 
neutrino mixing.  
In left-right symmetric models with seesaw (type-I), 
one expects this mixing to be of order 
$m_D/M_M \sim \sqrt{m_{\nu}/M_M}$, where $m_D$
is ($M_M$) the Dirac mass (Majorana mass) for the (right-handed)
neutrinos and $m_{\nu}$ is the light neutrino mass. 
This, in general,
is expected to be a small number of order $\sum_{j=1}^3 U_{ej}V_{ej}
\sim 10^{-5} \sqrt{(\frac{m_{\nu}}{\rm 0.1 eV})(\frac{\rm 1
 TeV}{M_M})}$.  
In this case one expects the mass mechanism to
dominate over both $\langle\lambda\rangle$ and $\langle\eta\rangle$,
given current limits on $W_L-W_R$ mixing \cite{Beringer:1900zz} and
lower limits on the $W_R$ mass from 
LHC~\cite{Khachatryan:2014dka,Aad:2015xaa}.  
However, as in the LQ example
model discussed previously in section \ref{subsect:tree},
contributions to the neutrino mass matrix contain a sum over the three
heavy right-handed neutrinos. In the case of severe fine-tuning of the
parameters entering the neutrino mass matrix, the connection between
the light-heavy neutrino mixing and $\meff$ can be avoided, see
section \ref{subsect:tree}. 
In this particular part of parameter
space, the incomplete $\sum_{j=1}^3 U_{ej}V_{ej}$ could in principle
be larger than the naive expectation.  
Recall that the current bound
on non-unitarity of $U$ is of the order of 1 \%
\cite{Escrihuela:2015wra}. 
For $\sum_{j=1}^3 U_{ej}V_{ej}$ as large as
$\sum_{j=1}^3 U_{ej}V_{ej} \sim {\cal O}(10^{-2})$
$\langle\lambda\rangle$ and/or $\langle\eta\rangle$ could dominate
over the mass mechanism, even after taking into account all other
existing limits. We stress again that this is not the natural
expectation.

In summary, there are some particular decompositions of $d=9$
operators containing the SM W or Higgs boson. 
In those cases the $d=9$ operator scales as 
$1/(\Lambda^3 v_{\rm SM}^2)$ and can be 
as important as the corresponding decomposition of the $d=7$
operator.

\section{Summary}
\label{sect:sum}

We have studied $d=7$ $\Delta L=2$ operators and their relation with
the long-range part of the amplitude for $\znbb$ decay. We have given
the complete list of decompositions for the relevant operators and
discussed a classification scheme for these decompositions based on
the level of perturbation theory, at which the different models
produce neutrino masses. For tree-level and 1-looop neutrino mass
models we expect that the mass mechanism is more important than the
long-range ($\pslash$-enhanced) amplitude. We have discussed how this
conclusion may be avoided in highly fine-tuned regions in parameter
space. For 2-loop neutrino mass models based on $d=7$ operators, the
long-range amplitude usually is more important than the mass
mechanism. To demonstrate this, we have discussed in some detail a
model based on ${\cal O}_8$.

We also discussed the connection of our work with previously 
considered long-range contributions in left-right symmetric models. 
This served to point out some particularities about the operator 
classification, that we rely on, in cases where higher order 
operators, such as $d=9$ (${\cal O}_9\propto \Lambda_{\rm LNV}^{-5}$),  
are effectively reduced to lower order operators, i.e. 
$d=7$ (${\cal O}_{9}^{\text{eff}} 
\propto \Lambda_{\rm LNV}^{-3}\times \Lambda_{\rm EW}^{-2} $).

Our main results are summarized in tabular form in the appendix, where
we give the complete list of possible models, which lead to
contributions to the long-range part of the amplitude for $\znbb$
decay. From this list one can deduce, which contractions can lead 
to interesting phenomenology, i.e. models that are testable also at the 
LHC.


\bigskip
\centerline{\bf Acknowledgements}

\medskip

M.H. thanks the Universidad Tecnica Federico Santa Maria, Valparaiso,
for hospitality during his stay. M.H. is supported by the Spanish
grants FPA2014-58183-P and Multidark CSD2009-00064 (MINECO), and
PROMETEOII/2014/084 (Generalitat Valenciana). J.C.H. is supported by
Fondecyt (Chile) under grants 11121557 and by CONICYT (Chile) project
791100017.
The research of T.O. is supported by JSPS Grants-in-Aid for Scientific
Research on Innovative Areas {\it Unification and Development of the
Neutrino Science Frontier} {\sf Number 2610 5503}.

\section{Appendix}
\label{app}

\begin{figure}[t]
 \unitlength=1cm
 \begin{picture}(6,4.5)
  \put(0,0){\includegraphics[width=6cm]{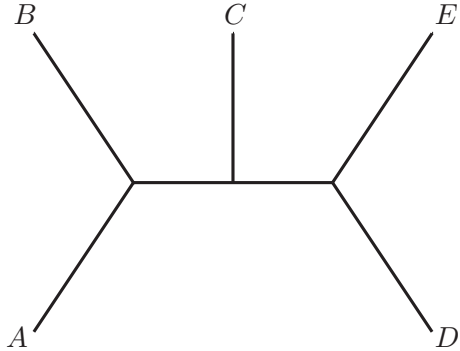}}
  \put(0,0){$A$}
  \put(0.1,4.3){$B$}
  \put(2.9,4.3){$C$}
  \put(5.7,0){$D$}
  \put(5.7,4.3){$E$}
 \end{picture}
 \caption{Topology for tree-level decompositions 
 of Babu-Leung operator \#3, \#4, and \#8.
 Once the SM fields, $A,B,C,D$, and $E$, 
 are assigned to the outer legs, 
 the Lorentz nature and the SM gauge charges 
 of the mediation fields are 
 uniquely determined.
 The assignments of the outer fields 
 are expressed as $(AB)(C)(DE)$
 and listed at the ``Decompositions'' column
 in Tabs.~\ref{Tab:BL3}-\ref{Tab:BL8}.}
\label{Fig:BL-decom}
\end{figure}

Here we present the summary tables of all tree-level
decompositions of the Babu-Leung operators \#3 (Tab.~\ref{Tab:BL3}),
\#4 (Tab.~\ref{Tab:BL4}), and \#8 (Tab.~\ref{Tab:BL8}) with mass
dimension $d=7$.  The effective operators are decomposed into 
renormalizable interactions by assigning the fields to the outer legs
of the tree diagram shown in Fig.~\ref{Fig:BL-decom}.
The assignments of the outer fields are shown at the
``Decompositions'' column, and the (inner) fields required by the
corresponding decompositions are listed at the ``Mediators'' column.
The symbols $S$ and $\psi$ represents the Lorentz nature of the
mediators: $S^{(')}$ is a scalar field, 
and $\psi_{L(R)}$ is a left(right)-handed fermion.  
The
charges of the mediators under the SM gauge groups are identified and
expressed with the format $(SU(3)_{c}, SU(2)_{L})_{U(1)_{Y}}$.  It is
easy to find the contributions of the effective operators to
neutrinoless double beta decay processes at the ``Projection to the
basis ops.'' column.  The basis operators are defined as
\begin{align}
 \mathcal{O}_{3a}(\alpha, \beta)
 \equiv&
 ({\rm i}\tau^{2})^{ij}
 ({\rm i}\tau^{2})^{kl}
 \{
 (\overline{L^{c}}_{\alpha})_{i}^{a}
 (L_{\beta})_{ja}
 \}
 \{
 (\overline{d_{R}})^{I b}
 (Q)_{I k b}
 \}
 H_{l}
 \nonumber 
 \\
 \supset&
 \frac{1}{4}
 \left[
 j_{S+P}^{\dagger}(\alpha,\beta)
 -
 j_{S+P}^{\dagger}(\beta,\alpha)
 \right]
 J_{S+P}^{\dagger}
 H^{0},
 \allowbreak
 \\
 \mathcal{O}_{3b}(\alpha, \beta)
 \equiv&
 ({\rm i}\tau^{2})^{ik}
 ({\rm i}\tau^{2})^{jl}
 \{
 (\overline{L^{c}}_{\alpha})_{i}^{a}
 (L_{\beta})_{ja}
 \}
 \{
 (\overline{d_{R}})^{I b}
 (Q)_{I k b}
 \}
 H_{l}
 \nonumber 
 \\
 \supset
 &
 -
 \frac{1}{4}
 j_{S+P}^{\dagger}(\beta,\alpha)
 J_{S+P}^{\dagger}
 H^{0}
 +
 \frac{1}{4}
 \{
 \overline{
 \ell^{c}
 }_{\alpha}
 (1-\gamma^{5})
 \ell_{\beta}
 \}
 J_{S+P}^{\dagger}
 H^{+},
  \allowbreak
 \\
  \mathcal{O}_{3a}^{\text{ten.}}(\alpha, \beta)
 \equiv&
 ({\rm i}\tau^{2})^{ij}
 ({\rm i}\tau^{2})^{kl}
 \{
 (\overline{L^{c}}_{\alpha})_{i}^{a}
 {(\sigma^{\rho \sigma})_{a}}^{b}
 (L_{\beta})_{jb}
 \}
 \{
 (\overline{d_{R}})^{I c}
 {(\sigma_{\rho \sigma})_{c}}^{d}
 (Q)_{I k d}
 \}
 H_{l}
 \nonumber 
 \\
 \supset&
 -
 \frac{1}{16}
 \left[
 (j_{T_{R}}^{\dagger})^{\rho \sigma} (\alpha,\beta)
 +
 (j_{T_{R}}^{\dagger})^{\rho \sigma} (\beta,\alpha)
 \right]
 (J_{T_{R}}^{\dagger})_{\rho \sigma}
 H^{0},
 \allowbreak
 \\
 \mathcal{O}_{3b}^{\text{ten.}}(\alpha, \beta)
 \equiv&
 ({\rm i}\tau^{2})^{ik}
 ({\rm i}\tau^{2})^{jl}
 \{
 (\overline{L^{c}}_{\alpha})_{i}^{a}
 {(\sigma^{\rho \sigma})_{a}}^{b}
 (L_{\beta})_{jb}
 \}
 \{
 (\overline{d_{R}})^{I c}
 {(\sigma_{\rho \sigma})_{c}}^{d}
 (Q)_{I k d}
 \}
 H_{l}
 \nonumber 
 \\
 \supset&
  -
 \frac{1}{16}
 (j_{T_{R}}^{\dagger})^{\rho\sigma}
 (\beta,\alpha)
 (J_{T_{R}}^{\dagger})_{\rho \sigma}
 H^{0}
 -\frac{1}{16}
 \{
 \overline{{\ell}^{c}}_{\alpha}
 \gamma^{\rho \sigma} (1-\gamma^{5})
 \ell_{\beta}
 \}
 (J_{T_{R}}^{\dagger})_{\rho \sigma}
  H^{+},
 \allowbreak
 \\
 \mathcal{O}_{4a}(\alpha,\beta) \equiv& 
  ({\rm i} \tau)^{jk}
 (\overline{L^{c}}_{\alpha})_{i}^{a}
 (L_{\beta})_{ja}
 (\overline{Q})^{Ii}_{\dot{a}}
 (u_{R})_{I}^{\dot{a}}
 H_{k}
 \nonumber
 \\
 \supset&
 \frac{1}{4}
 j_{S+P}^{\dagger} (\beta,\alpha)
 J_{S-P}^{\dagger} H^{0}
 -
 \frac{1}{4}
 \overline{\ell^{c}}_{\alpha}
 (1-\gamma^{5}) \ell_{\beta}
 J_{S-P}^{\dagger} H^{+},
  \allowbreak
 \\
 \mathcal{O}_{4b}(\alpha,\beta) \equiv&
 ({\rm i} \tau^{2})^{ij}
 (\overline{L^{c}}_{\alpha})_{i}^{a}
 (L_{\beta})_{ja}
 (\overline{Q})^{Ik}_{\dot{a}}
 (u_{R})_{I}^{\dot{a}}
 H_{k}
 \nonumber 
 \\
 \supset&
 \frac{1}{4}
 \left[
 j_{S+P}^{\dagger} (\alpha,\beta)
 -
 j_{S+P}^{\dagger} (\beta,\alpha)
 \right]
 J_{S-P}^{\dagger} H^{0},
  \allowbreak
 \\
 \mathcal{O}_{8}(\alpha,\beta) \equiv& 
 (\overline{L^{c}}_{\alpha})_{i}^{a}
 (\sigma^{\rho})_{a \dot{a}}
 (e_{R \beta})^{\dot{a}}
 (\overline{d_{R}})^{I b}
 (\sigma_{\rho})_{b \dot{b}}
 (u_{R})_{I}^{\dot{b}}
 ({\rm i} \tau^{2})^{ij}
 H_{j}
 \nonumber 
 \\
 \supset&
 \frac{1}{4}
 (j_{V+A}^{\dagger})^{\rho} 
 (J_{V+A}^{\dagger})_{\rho}
 H^{0}
 -
 \frac{1}{4}
 \left\{
 \overline{\ell^{c}}_{\alpha} \gamma^{\rho} (1+\gamma^{5}) \ell_{\beta}
 \right\}
 (J_{V+A}^{\dagger})_{\rho} H^{+}.
\end{align}
%
%
Here we explicitly write all the indices:
$\alpha, \beta$ for lepton flavour,
the lower (upper) $I$ for $\boldsymbol{3}$ ($\boldsymbol{\bar{3}}$) 
of $SU(3)$ colour,
$i,j,k,l$ for $\boldsymbol{2}$ of $SU(2)$ left,
$\rho,\sigma$ for Lorentz vector, 
and
$a,b,c,d$ ($\dot{a},\dot{b}$) 
for left(right)-handed Lorentz spinor.
The lowest-loop 
contributions (i.e., dominant contributions) 
to neutrino masses 
are found at the columns ``$m_{\nu}$''.
We are mainly interested in decompositions (=proto-models) 
where new physics contributions to $\znbb$
can compete with the mass mechanism contribution 
mediated by the effective neutrino mass 
$\langle m_{\nu} \rangle$.  
%
An annotation ``w. (additional interaction)''
is given in the column of ``$m_{\nu}$@1loop''
for some decompositions.
This shows that one can draw the 1-loop diagram, 
putting the interactions that appear in the decomposition
and {\it the additional interaction} together.
The additional interactions given in the tables 
are not included in the decomposition but 
are not forbidden by the SM gauge symmetries, nor 
can they be eliminated by any (abelian) discrete symmetry, 
without removing at least some of the interactions present 
in the decomposition.
For example, 
using the interactions appear in 
decomposition \#11 of Babu-Leung operator \#8 
(see Tab.~\ref{Tab:BL8}),
one can construct two 2-loop neutrino mass diagrams
mediated by the Nambu-Goldstone boson $H^{+}$,
whose topologies are $T2^{B}_{2}$ and $T2^{B}_{4}$
of \cite{Sierra:2014rxa}.
This also corresponds to the 2-loop neutrino mass model 
labelled with $\mathcal{O}_{8}^{1}$ in \cite{Cai:2014kra}.
However, to regularize the divergence in diagram 
$T2^{B}_{4}$, the additional interaction 
$(\overline{Q^{c}})_{Ii}^{a} ({\rm i} \tau^{2})^{ij} (L)_{ia} S^{I}$
is necessary, 
and this interaction generates a 1-loop neutrino mass diagram.
Consequently, this decomposition should be regarded as a 1-loop 
neutrino mass model.\footnote{%
We note that the same argument holds for 
all decompositions containing the scalar $S_{{\bar 3},1,1/3}$ 
listed in \cite{Helo:2015fba} as 2-loop $d=7$ models.}
We also show the 1-loop neutrino mass models that require 
an additional interaction with an additional field 
(second Higgs doublet $H'$)
with bracket.\footnote{%
Although the interaction 
$(\partial_{\rho} H)_{i} ({\rm i} \tau^{2})^{ij} H_{j} V^{\rho}$
listed in Tab.~\ref{Tab:BL8} can be constructed only with 
the SM Higgs doublets $H$ and the vector mediator $V$ of the $d=7$
operator,
the interaction does not appear in the models where  
the vector mediator $V$ is the gauge boson of an extra gauge symmetry.
However, if we allow the introduction of an additional Higgs doublet
$H'$, we can have the $(\partial_{\rho} H)_{i} H'^{\dagger i} V^{\rho}$
through the mixing between $H$ and $H'$.
}

%
The two contributions to $\znbb$ are compared in Sec.~\ref{sect:class} 
with some concrete examples.
The comparison is summarized at Tab.~\ref{Tab:Comparison}.
\begin{table}[t]
 \begin{tabular}{ccccc}
  \hline \hline 
  Eff. op. 
  & Decom. 
  & $(m_{\nu})_{\alpha\beta}$ 
  & \begin{minipage}{3cm}
     $\Lambda_{7}$ [GeV] 
     \vspace{-0.2cm}\\
     suggested 
     \vspace{-0.2cm}\\
     by $m_{\nu}=0.05$ eV
    \end{minipage}
  &
  $\mathcal{A}^{\rm MM}/\mathcal{A}^{\rm LR}$
  \\
  \hline
  $LL \overline{d_{R}} Q H$ 
  & \#1,3,4,5,6,9
      & $\frac{v^{2}}{\Lambda}$
	  & $\sim 10^{15}$
	      & $\frac{\Lambda^{2}}{p_{F} v} \sim 10^{28}$
	  \\
  & \#2,7,8
      & $\frac{y_{b}}{16\pi^{2}}\frac{v^{2}}{\Lambda}$
	  & $\sim 10^{11}$
	      & 
  $\frac{y_{b}}{16 \pi^{2}}
  \frac{\Lambda^{2}}{p_{F} v} \sim 10^{17}$
      \\
  & $\biggl[\#1,3,5,8$ 
      & $\frac{y_{b} g^{2}}{(16\pi^{2})^{2}} \frac{v^{2}}{\Lambda}$
	  $(\alpha\neq\beta)$
	  & $\sim 10^{8}$
	         & 
  $\frac{y_{b} g^{2}}{(16 \pi^{2})^{2}}
  \frac{\Lambda^{2}}{p_{F} v} \sim 10^{9} \biggr]$
  \vspace{0.1cm}
  \\
  \hline
  $LL \overline{Q} u_{R} H$
  & \#1,3,4,5,6,8,9
      & $\frac{v^{2}}{\Lambda}$
	  & $\sim 10^{15}$
	      & $\frac{\Lambda^{2}}{p_{F} v} \sim 10^{28}$
  \\
  & \#2,7,8
      & $\frac{y_{t}}{16 \pi^{2}} \frac{v^{2}}{\Lambda}$
	  & $\sim 10^{12}$
	  & $\frac{y_{t}}{16 \pi^{2}} \frac{\Lambda^{2}}{p_{F} v} \sim 10^{21}$
	      \\
  & $\biggl[
      \#1,3,5,8$
      & $\frac{y_{t} g^{2}}{(16\pi^{2})^{2}} \frac{v^{2}}{\Lambda}$
	  $(\alpha \neq \beta)$
	  & $\sim 10^{10}$
	      & $\frac{y_{t} g^{2}}{(16 \pi^{2})^{2}}
  \frac{\Lambda^{2}}{p_{F} v} \sim 10^{14}\biggr]$
  \vspace{0.1cm}
  \\
  \hline
  $Le_{R} \overline{d_{R}} u_{R} H$
  & 
      \#5,8,14
      & $\frac{v^{2}}{\Lambda}$
	  & $\sim 10^{15}$ 
	      & $\frac{\Lambda^{2}}{p_{F} v} \sim 10^{28}$
  \\
  & \#2,12
      & $\frac{y_{t}}{16 \pi^{2}} \frac{v^{2}}{\Lambda}$
	  & $\sim 10^{12}$
      & $\frac{y_{t}}{16 \pi^{2}} \frac{\Lambda^{2}}{p_{F} v} \sim 10^{21}$
      \\
  & \#3,11
      & $\frac{y_{b}}{16\pi^{2}}\frac{v^{2}}{\Lambda}$
	  & $\sim 10^{11}$
	      & 
  $\frac{y_{b}}{16 \pi^{2}}
  \frac{\Lambda^{2}}{p_{F} v} \sim 10^{17}$
  \\
  & \begin{minipage}{3cm}
      \#1,4,6,7,9,
     \vspace{-0.15cm}\\
     10,13,15
    \end{minipage}
  & 
      $ y_{\ell_{\beta}} \frac{y_{b} y_{t} g^{2}}{(16 \pi^{2})^{2}} 
	  \frac{v^{4}}{\Lambda^{3}}$
	  &
	  $\sim 10^{3}$ 
	  ($\beta=\tau$)
	  & 
	      $ 
	      y_{e} 
	      \frac{y_{b} y_{t} g^{2}}{(16 \pi^{2})^{2}}
	      \frac{v}{p_{F}} 
	      \sim 10^{-9}$ 
	      \\
  \hline
  \hline
 \end{tabular}
\caption{Comparison between the amplitude $\mathcal{A}^{\rm LR}$ of
  new physics long-range contributions to $\znbb$ and that
  $\mathcal{A}^{\rm MM}$ of the mass mechanism. When the neutrino mass
  is generated at the tree and one-loop level, the new physics scale
  $\Lambda_{7}$ must be sufficiently high to reproduce the correct
  size of neutrino masses, consequently, the long-range contributions
  $\mathcal{A}^{\rm LR}$ are suppressed and the mass mechanism
  dominates the contribution to $\znbb$. As usual in such operator
  analysis, these estimates do not take into account that some non-SM
  Yukawa couplings, appearing in the ultra-violet completion of the
  operators, could be sizably smaller than one, which would lead to
  lower scales $\Lambda_7$. Also, for loop model the scales could be
  overestimated, since they neglect loop integrals.  The neutrino
  masses generated at the two-loop level from the decompositions of
  the Babu-Leung \#8 operator should be estimated with $d=7$
  $LLHHHH^{\dagger}$ operator (as illustrated in
  sect. \ref{subsect:2lp}).  In addition, they receive additional
  suppression from the lepton Yukawa coupling $y_{\ell_{\beta}}$,
  which further lowers the new physics scale $\Lambda_{7}$.  Note that
  in particular for the 2-loop $d=7$ models, as the concrete example
  in sect. \ref{subsect:2lp} shows, the estimate for $\mathcal{A}^{\rm
    MM}/\mathcal{A}^{\rm LR}$  can vary by several orders of
    magnitude, depending on parameters.  However, both the estimate
    shown here and the explicit calculation in sect. \ref{subsect:2lp}
    give numbers $\mathcal{A}^{\rm MM}/\mathcal{A}^{\rm LR} \ll 1$ ,
    such that the long-range contribution dominates always over the mass
    mechanism for these decompositions. }
\label{Tab:Comparison}
\end{table}
In short, the mass mechanism dominates $\znbb$ 
if neutrino masses are generated at the tree or the 1-loop level.
When neutrino masses are generated from 2-loop diagrams,
new physics contributions to $\znbb$ become comparable with 
the mass mechanism contribution and can be large enough 
to be within reach of  
the sensitivities of next generation experiments.
However,
the 2-loop neutrino masses generated from the decompositions 
of the Babu-Leung operators of \#3 and \#4 are 
anti-symmetric with respect to the flavour indices,
such as the original Zee model and, thus, are already excluded 
by oscillation experiments.
Therefore, if we adopt those decompositions as neutrino mass models,
we must extend the models to make the neutrino masses
compatible with oscillation data.
In such models, the extension part 
controls the mass mechanism contribution and 
also the new physics contribution to $\znbb$,
and consequently, we cannot compare the contributions 
without a full description of the models including 
the extension.
Nonetheless, it might be interesting to point out that
decomposition \#8 of the Babu-Leung \#3
contains the tensor operator $\mathcal{O}_{3a}^{\text{ten.}}(e,e)$,
which gives a contribution to $\znbb$ 
and generates neutrino masses with the $(e,e)$ component 
at the two-loop level.
On the other hand, 
2-loop neutrino mass models inspired by 
decompositions of Babu-Leung \#8 possess
a favourable flavour structure.
This possibility has been investigated 
in Sec.~\ref{subsect:2lp}
with a concrete example.

\begin{table}[t]
\hspace*{-1.5cm}
\begin{tabular}{cccccccc}
\hline \hline
 \# & Decompositions & \multicolumn{2}{c}{Mediators} 
 & Projection to the basis ops.
 & $m_{\nu}$@tree
 & $m_{\nu}$@1loop
 & $m_{\nu}$@2loop
 \\
 \hline
 \#1 
 & $(L_{\alpha}L_{\beta})(H)(\overline{d_{R}}Q)$ 
     & $S({\bf 1},{\bf 1})_{+1}$ 
	 & $S'({\bf 1},{\bf 2})_{+\frac{1}{2}}$
	     & $-\mathcal{O}_{3a}(\alpha,\beta)$
		 & ---
		     & 
			 \begin{minipage}{2cm}
			  \vspace{0.1cm}
			  T$\nu$I-ii
			  \\
			  w.$\overline{\ell_{R}}LS'^{\dagger}$
			  \vspace{0.1cm}
			 \end{minipage}
 &  
     \begin{minipage}{2cm}
      \vspace{0.1cm}
     $T2^{\rm B}_{4} (\alpha \neq \beta)$
      \\
      $\mathcal{O}_{3}^{7}$ in \cite{Cai:2014kra}
     \end{minipage}

 \\
 \cline{3-8}
 & 
 & $S({\bf 1},{\bf 3})_{+1}$ 
	 & $S'({\bf 1},{\bf 2})_{+\frac{1}{2}}$
	     & 
	$-\mathcal{O}_{3b}(\alpha,\beta) -\mathcal{O}_{3b}(\beta,\alpha)$
		 & 
		    type II
 & 
     &
\\
\hline
 \#2
 & $(L_{\alpha}Q)(H)(\overline{d_{R}}L_{\beta})$
     & $S(\overline{\bf 3}, {\bf 1})_{+\frac{1}{3}}$
	 & $S'(\overline{\bf 3}, {\bf 2})_{-\frac{1}{6}}$
	     & 
		 $\frac{1}{2} \mathcal{O}_{3b}(\alpha,\beta)-
		 \frac{1}{2}
		 \mathcal{O}_{3b}^{\text{ten.}}(\alpha,\beta) $
		 &
		     ---
		     & 
		 \begin{minipage}{2cm}
		  \vspace{0.1cm}
		  T$\nu$I-ii
		  \cite{AristizabalSierra:2007nf}
		  \\
		  $\mathcal{O}_{3}^{8}$ in \cite{Cai:2014kra}
		 \end{minipage}
 &
     \cite{Babu:2001ex,Babu:2010vp}
 \\
 \cline{3-8}
&
     & $S(\overline{\bf 3}, {\bf 3})_{+\frac{1}{3}}$
	 & $S'(\overline{\bf 3}, {\bf 2})_{-\frac{1}{6}}$
	     &\begin{minipage}{4.8cm}
	       \vspace{0.1cm}
	       $\frac{1}{2} \mathcal{O}_{3a}(\alpha,\beta)
	       -\frac{1}{2} \mathcal{O}_{3a}^{\text{ten.}}(\alpha,\beta)
	       $
	       \\
	       $
	       -
	       \frac{1}{2}
	       \mathcal{O}_{3b} (\beta,\alpha) 
	       -
	       \frac{1}{2}
	       \mathcal{O}_{3b}^{\text{ten.}} (\beta,\alpha)$
	       \vspace{0.1cm}
	      \end{minipage}
 & ---
     & 
 \begin{minipage}{2cm}
  \vspace{0.1cm}
  T$\nu$I-ii
	\cite{AristizabalSierra:2007nf}
  \\
  $\mathcal{O}_{3}^{9}$ in \cite{Cai:2014kra}
 \end{minipage}
 &
     \cite{Babu:2001ex}
\\
\hline
 \#3 &
     $(L_{\alpha}L_{\beta})(Q)(\overline{d_{R}}H)$
     &
	 $S({\bf 1},{\bf 1})_{+1}$
	 & $\psi_{L,R}({\bf 3},{\bf 2})_{-\frac{5}{6}}$
	     &$-\mathcal{O}_{3a}(\alpha,\beta)$
		 & ---
		     & 
			 \begin{minipage}{2cm}
			  \vspace{0.1cm}
			  $\left[
			  \begin{minipage}{1.7cm}
			  T$\nu$I-ii
			  \\
			  w.$S^{\dagger}HH'$
			  \end{minipage}
			   \right]$
			 \end{minipage}
			 & 
 \begin{minipage}{2cm}
  \vspace{0.1cm}
  $T2^{\rm B}_{1} (\alpha\neq \beta)$
  \\
  $\mathcal{O}_{3}^{1}$ in \cite{Cai:2014kra}
 \end{minipage}
 \\
 \cline{3-8}
 &
     &
	 $S({\bf 1},{\bf 3})_{+1}$
	 &
	     $\psi_{L,R}({\bf 3},{\bf 2})_{-\frac{5}{6}}$
	     &
		 $-\mathcal{O}_{3b}(\alpha,\beta)
		 - \mathcal{O}_{3b}(\beta,\alpha)$
		 &
		      type II
\\
\hline
 \#4 & $(L_{\alpha} H)(Q)(\overline{d_{R}}L_{\beta})$
     & $\psi_{R}({\bf 1},{\bf 1})_{0}$
	 & $S(\overline{\bf 3},{\bf 2})_{-\frac{1}{6}}$
	     &
		 $\frac{1}{2} \mathcal{O}_{3b}(\beta,\alpha)
		 +
		 \frac{1}{2}
		 \mathcal{O}_{3b}^{\text{ten.}}(\beta,\alpha)$
		 &
		      type I
 \\
 \cline{3-8}
 &
     & $\psi_{R}({\bf 1},{\bf 3})_{0}$
	 & $S(\overline{\bf 3},{\bf 2})_{-\frac{1}{6}}$
	     &
		 \begin{minipage}{4.8cm}
		  \vspace{0.1cm}
		  $-\frac{1}{2} \mathcal{O}_{3a}(\alpha,\beta)
		  +\frac{1}{2} \mathcal{O}_{3a}^{\text{ten.}}(\alpha,\beta)
		  $
		  \\
		  $
		  -
		  \frac{1}{2}
		  \mathcal{O}_{3b} (\alpha,\beta) 
		  +
		  \frac{1}{2}
		  \mathcal{O}_{3b}^{\text{ten.}} (\alpha,\beta) $
		  \vspace{0.1cm}
		 \end{minipage}
 &
		      type III
\\
\hline
\#5 & $(L_{\alpha}L_{\beta})(\overline{d_{R}})(QH)$
     & $S({\bf 1},{\bf 1})_{+1}$ 
	 & $\psi_{L,R}({\bf 3},{\bf 1})_{+\frac{2}{3}}$
	     & $\mathcal{O}_{3a}(\alpha,\beta)$
		 & ---
		     & 			 
			 \begin{minipage}{2cm}
			  \vspace{0.1cm}
			  $\left[
			  \begin{minipage}{1.7cm}
			  T$\nu$I-ii
			  \\
			  w.$S^{\dagger}HH'$
			  \end{minipage}
			  \right]$
			 \end{minipage}
			 & 
  \begin{minipage}{2cm}
	\vspace{0.1cm}
   $T2^{\rm B}_{2} (\alpha \neq \beta)$
  \\
   $\mathcal{O}_{3}^{2}$ in \cite{Cai:2014kra}
 \end{minipage}
 \\
 \cline{3-8}
 &
      & $S({\bf 1},{\bf 3})_{+1}$ 
	 & $\psi_{L,R}({\bf 3},{\bf 3})_{+\frac{2}{3}}$
	     & $-\mathcal{O}_{3b}(\alpha,\beta) 
	     -\mathcal{O}_{3b}(\beta,\alpha)$
	     & 
		      type II
\\
\hline
\#6 & $(L_{\alpha}Q)(\overline{d_{R}})(L_{\beta}H)$
     & $S(\overline{\bf 3},{\bf 1})_{+\frac{1}{3}}$ 
	 & $\psi_{R}({\bf 1},{\bf 1})_{0}$
	     & $-\frac{1}{2}\mathcal{O}_{3b}(\alpha,\beta)
		 +\frac{1}{2}\mathcal{O}_{3b}^{\text{ten.}}(\alpha,\beta) $
		 & 
		      type I
 \\
 \cline{3-8}
 &
     & $S(\overline{\bf 3},{\bf 3})_{+\frac{1}{3}}$ 
	 & $\psi_{R}({\bf 1},{\bf 3})_{0}$
	     & 
		 \begin{minipage}{4.8cm}
		  \vspace{0.1cm}
		  $
		  \frac{1}{2} \mathcal{O}_{3a}(\alpha,\beta)
		  -
		  \frac{1}{2} \mathcal{O}_{3a}^{\text{ten.}}(\alpha,\beta)
		  $
		  \\
		  $
		  -
		  \frac{1}{2}
		  \mathcal{O}_{3b} (\beta,\alpha) 
		  -
		  \frac{1}{2}
		  \mathcal{O}_{3b}^{\text{ten.}} (\beta,\alpha)$
		  \vspace{0.1cm}
		 \end{minipage}
		 &  
      type III
\\
\hline
\#7 & $(L_{\alpha}Q)(L_{\beta})(\overline{d_{R}}H)$
     & $S(\overline{\bf 3},{\bf 1})_{+\frac{1}{3}}$ 
	 & $\psi_{L,R}({\bf 3},{\bf 2})_{-\frac{5}{6}}$
	     & $\frac{1}{2} \mathcal{O}_{3b}(\alpha,\beta)
		 -
		 \frac{1}{2}
		 \mathcal{O}_{3b}^{\text{ten.}}(\alpha,\beta)$
		 & ---
		     &
			 \begin{minipage}{2cm}
			  \vspace{0.1cm}
			  T$\nu$I-iii
			  \\
			  $\mathcal{O}_{3}^{4}$ in \cite{Cai:2014kra}
			  	  \vspace{0.1cm}
			 \end{minipage}  
 \\
 \cline{3-8}
 &
     & $S(\overline{\bf 3},{\bf 3})_{+\frac{1}{3}}$ 
	 & $\psi_{L,R}({\bf 3},{\bf 2})_{-\frac{5}{6}}$
	     & 
		 \begin{minipage}{4.8cm}
		  \vspace{0.1cm}
		  $
		  \frac{1}{2} \mathcal{O}_{3a}(\alpha,\beta)
		  -
		  \frac{1}{2} \mathcal{O}_{3a}^{\text{ten.}}(\alpha,\beta)
		  $
		  \\
		  $
		  -
		  \frac{1}{2}
		  \mathcal{O}_{3b} (\beta,\alpha) 
		  -
		  \frac{1}{2}
		  \mathcal{O}_{3b}^{\text{ten.}} (\beta,\alpha)$
		  \vspace{0.1cm}
		 \end{minipage}
		 & ---
		     &
			 \begin{minipage}{2cm}
			  \vspace{0.1cm}
			  T$\nu$I-iii
			  \\
			  $\mathcal{O}_{3}^{5}$ in \cite{Cai:2014kra}
			 \end{minipage}
&
\\
\hline
\#8 & $(\overline{d_{R}}L_{\alpha})(L_{\beta})(QH)$
     & $S({\bf 3},{\bf 2})_{+\frac{1}{6}}$
	 & $\psi_{L,R}({\bf 3},{\bf 1})_{+\frac{2}{3}}$
	     &
		 $-\frac{1}{2} \mathcal{O}_{3a}(\alpha,\beta)
		 -\frac{1}{2}
		 \mathcal{O}_{3a}^{\text{ten.}}(\alpha,\beta)$
		 &
		     ---
		     & ---
			 &
\begin{minipage}{2cm}
 \vspace{0.1cm}
  $T2^{\rm B}_{2} (m_{\nu})_{\alpha \neq \beta}$
  \\
 $\mathcal{O}_{3}^{3}$ in \cite{Cai:2014kra},
 \cite{Babu:2011vb} 
  \vspace{0.1cm}
\end{minipage}
 \\
 \cline{3-8}
 & & $S({\bf 3},{\bf 2})_{+\frac{1}{6}}$
     & $\psi_{L,R}({\bf 3},{\bf 3})_{+\frac{2}{3}}$
	 &
		 \begin{minipage}{4.8cm}
		  \vspace{0.1cm}
		  $
		  \frac{1}{2} \mathcal{O}_{3b}(\alpha,\beta)
		  +
		  \frac{1}{2} \mathcal{O}_{3b}^{\text{ten.}}(\alpha,\beta)
		  $
		  \\
		  $
		  +
		  \frac{1}{2}
		  \mathcal{O}_{3b} (\beta,\alpha) 
		  -
		  \frac{1}{2}
		  \mathcal{O}_{3b}^{\text{ten.}} (\beta,\alpha)$
		  \vspace{0.1cm}
		 \end{minipage}
 &
     ---
     & 
	 \begin{minipage}{2cm}
	  \vspace{0.1cm}
	  T$\nu$I-iii
	  \\
	  $\mathcal{O}_{3}^{6}$ in \cite{Cai:2014kra}
	 \end{minipage}
 & 
\\
\hline
\#9 & $(L_{\alpha}H)(L_{\beta})(\overline{d_{R}}Q)$
     & $\psi_{R}({\bf 1},{\bf 1})_{0}$
	 & $S({\bf 1},{\bf 2})_{+\frac{1}{2}}$
	     &
		 $\mathcal{O}_{3b}(\beta,\alpha)$
		 & 
		      type I
 \\
 \cline{3-8}
 &
     & $\psi_{R}({\bf 1},{\bf 3})_{0}$
	 & $S({\bf 1},{\bf 2})_{+\frac{1}{2}}$
	     & 
		 $\mathcal{O}_{3a}(\alpha,\beta) 
		 + \mathcal{O}_{3b}(\alpha,\beta)$
		 & 
		      type III
	     \\
\hline \hline
\end{tabular}
\caption{Decompositions and projections of 
the $LL\overline{d_{R}}QH$ operator.
New physics contributions 
to $0\nu2\beta$ are given as 
the combinations of the basis operators 
in the ``Projection to the basis ops.'' column.
The tensor operators $\mathcal{O}^{\text{ten.}}$ 
play an important role 
in the long-range contribution.
The long-range contribution in R-parity violating SUSY models
corresponds to decomposition \#2~\cite{Babu:1995vh,Pas:1998nn}.
}
\label{Tab:BL3}
\end{table}

\begin{table}[t]
 \hspace*{-1.5cm}
\begin{tabular}{cccccccc}
\hline \hline
 \# & Decompositions & \multicolumn{2}{c}{Mediators} 
 & Projection to the basis ops.
 & $m_{\nu}$@tree
 & $m_{\nu}$@1loop
 & $m_{\nu}$@2loop
 \\
 \hline
 \#1 
 & $(L_{\alpha}L_{\beta})(H)(\overline{Q}u_{R})$
     & $S({\bf 1},{\bf 1})_{+1}$
	 & $S'({\bf 1},{\bf 2})_{+\frac{1}{2}}$
	     & $-\mathcal{O}_{4b}(\alpha,\beta)$
		 & ---
		     & 
			 \begin{minipage}{2cm}
			  \vspace{0.1cm}
			  T$\nu$I-ii
			  \\
			  w.$\overline{e_{R}}LS'^{\dagger}$
			  \vspace{0.1cm}
			 \end{minipage}
 &
 \begin{minipage}{2cm}
  \vspace{0.1cm}
$T2^{\rm B}_{4} (\alpha \neq \beta)$
      \\
$\mathcal{O}_{4}^{3}$ in \cite{Cai:2014kra}
     \end{minipage}
 \\
 \cline{3-8}
 &
     & $S({\bf 1},{\bf 3})_{+1}$
	 & $S'({\bf 1},{\bf 2})_{+\frac{1}{2}}$
	     &
		 $\mathcal{O}_{4a}(\alpha,\beta)+\mathcal{O}_{4a}(\beta,\alpha)$
		 & 
		    type II
	 \\
 \hline
 \#2
 & $(\overline{Q} L_{\alpha}) (H) (L_{\beta} u_{R})$
     & $V({\bf 3}, {\bf 1})_{+\frac{2}{3}}$
	  & $V'({\bf 3}, {\bf 2})_{+\frac{1}{6}}$
	     & $2 \mathcal{O}_{4a} (\alpha,\beta)$
		 & ---
		     & 
			T$\nu$I-ii
 \\
 \cline{3-8}
 &
     & $V({\bf 3}, {\bf 3})_{+\frac{2}{3}}$
	 & $V'({\bf 3}, {\bf 2})_{+\frac{1}{6}}$
	     & $2 \mathcal{O}_{4a} (\beta,\alpha) 
		 - 2 \mathcal{O}_{4b} (\alpha,\beta)$
		 & ---
		     & 
			T$\nu$I-ii
		 \\
 \hline
 \#3 
 & $(L_{\alpha} L_{\beta}) (\overline{Q}) (u_{R} H)$
     & $S({\bf 1}, {\bf 1})_{+1}$
	 & $\psi_{L,R}({\bf 3},{\bf 2})_{+\frac{7}{6}}$
	     & $\mathcal{O}_{4b} (\alpha,\beta)$
		 & ---
		     & \begin{minipage}{1.8cm}
			 \vspace{0.1cm}
			 $\left[
			 \begin{minipage}{1.7cm}
			  T$\nu$I-ii
			  \\
			  w.$S^{\dagger} HH'$
			 \end{minipage}
 \right]$
		       \end{minipage}
 &
 \begin{minipage}{2cm}
  \vspace{0.1cm}
 $T2^{\rm B}_{1} (\alpha \neq \beta)$
  \\
  $\mathcal{O}_{4}^{3}$ in \cite{Cai:2014kra}
 \end{minipage}
     \\
 \cline{3-8}
 &
     & $S({\bf 1}, {\bf 3})_{+1}$
	 & $\psi_{L,R}({\bf 3},{\bf 2})_{+\frac{7}{6}}$
	     & $\mathcal{O}_{4a} (\alpha,\beta) + 
		 \mathcal{O}_{4a} (\beta,\alpha)$
		 & 
		    type II
     \\
 \hline
 \#4
 & $(L_{\alpha} H)(\overline{Q})(L_{\beta} u_{R})$
     & $\psi_{R}({\bf 1}, {\bf 1})_{0}$
	 & $V({\bf 3}, {\bf 2})_{+\frac{1}{6}}$
	     & $2\mathcal{O}_{4a} (\beta,\alpha)$
		 & 
		    type I
     \\
 \cline{3-8}
 &
     & $\psi_{R}({\bf 1}, {\bf 3})_{0}$
	 & $V({\bf 3}, {\bf 2})_{+\frac{1}{6}}$
	     & $-2\mathcal{O}_{4a} (\alpha,\beta)
		 +2\mathcal{O}_{4b} (\alpha,\beta)$
		 &
		    type III
		 \\
 \hline
 \#5
  & $(L_{\alpha} L_{\beta})(u_{R})(\overline{Q} H)$
     & $S({\bf 1}, {\bf 1})_{+1}$
	 & $\psi_{L,R}(\overline{\bf 3}, {\bf 1})_{+\frac{1}{3}}$
	     & $\mathcal{O}_{4b} (\alpha,\beta)$
		 & ---
		     & 	
			 \begin{minipage}{2cm}
			  \vspace{0.1cm}
			  $\left[
			  \begin{minipage}{1.7cm}
			  T$\nu$I-ii
			  \\
			  w.$S^{\dagger} HH'$
			  \end{minipage}
			  \right]$
			 \end{minipage}
			 &  
 \begin{minipage}{2cm}
   \vspace{0.1cm}
 $T2^{\rm B}_{2} (\alpha \neq \beta)$
   \\
  $\mathcal{O}_{4}^{2}$ in \cite{Cai:2014kra} 
 \end{minipage}
     \\
 \cline{3-8}
 & & $S({\bf 1}, {\bf 3})_{+1}$
	 & $\psi_{L,R}(\overline{\bf 3}, {\bf 3})_{+\frac{1}{3}}$
	     & $\mathcal{O}_{4a} (\alpha,\beta) + 
		 \mathcal{O}_{4a} (\beta,\alpha)$
		 &
		    type II
     \\
 \hline
 \#6
  & $(\overline{Q} L_{\alpha}) (u_{R}) (L_{\beta}H)$
     &$ V({\bf 3},{\bf 1})_{+\frac{2}{3}}$
	 &$ \psi_{R}({\bf 1},{\bf 1})_{0}$
	     & $2\mathcal{O}_{4a}(\alpha,\beta)$
		 &
		    type I
		 \\
 \cline{3-8}
 & & $ V({\bf 3},{\bf 3})_{+\frac{2}{3}}$
	 &$ \psi_{R}({\bf 1},{\bf 3})_{0}$
	     & $-2\mathcal{O}_{4b}(\alpha,\beta)
		 -2\mathcal{O}_{4a}(\beta,\alpha)$
		 &
		    type III
	 \\
 \hline
 \#7 & $(\overline{Q} L_{\alpha}) (L_{\beta}) (u_{R} H)$
     & $V({\bf 3}, {\bf 1})_{+\frac{2}{3}}$
	 & $\psi_{L,R}({\bf 3},{\bf 2})_{+\frac{7}{6}}$
	     & $-2 \mathcal{O}_{4a}(\alpha,\beta)$
		 & ---
		     & 
			 T$\nu$I-iii

 \\
 \cline{3-8}
 & & $ V({\bf 3},{\bf 3})_{+\frac{2}{3}}$
	 & $\psi_{L,R}({\bf 3},{\bf 2})_{+\frac{7}{6}}$
	     & $2 \mathcal{O}_{4b}(\alpha,\beta) 
		 + 2 \mathcal{O}_{4a}(\beta,\alpha)$
		 & ---
		     & 
			 T$\nu$I-iii
		 \\
 \hline
 \#8 & $(L u_{R}) (L) (\overline{Q} H)$
     & $V(\overline{\bf 3},{\bf 2})_{-\frac{1}{6}}$
	 & $\psi_{L,R}({\bf 3},{\bf 1})_{-\frac{1}{3}}$
	     & $2 \mathcal{O}_{4b} (\alpha,\beta)$
		 & ---
		     & ---
			 & 
 $T2^{\rm B}_{2}$ ($\alpha \neq \beta$)
	     \\
 \cline{3-8}
 &&  $V(\overline{\bf 3},{\bf 2})_{-\frac{1}{6}}$
	 & $\psi_{L,R}({\bf 3},{\bf 3})_{-\frac{1}{3}}$
	     & $2 \mathcal{O}_{4a} (\alpha,\beta)
		 +2 \mathcal{O}_{4a} (\beta,\alpha)$
		 & ---
		     & 
			 T$\nu$I-iii
		 \\
 \hline
 \#9 & $(LH)(L)(\overline{Q} u_{R})$
     & $\psi_{R}({\bf 1},{\bf 1})_{0}$
	 & $S({\bf 1},{\bf 2})_{+\frac{1}{2}}$
	     & $- \mathcal{O}_{4a}(\beta,\alpha)$
		 &
		    type I
	 \\
 \cline{3-8}
 &&  $\psi_{R}({\bf 1},{\bf 3})_{0}$
	 & $S({\bf 1},{\bf 2})_{+\frac{1}{2}}$
	     & $-\mathcal{O}_{4a}(\alpha,\beta) 
		 + \mathcal{O}_{4b}(\alpha,\beta)$
		 &
		    type III
		 \\
 \hline \hline
\end{tabular}
\caption{Decomposition and projection of the $LL\overline{Q}u_{R}H$
 operator.}
\label{Tab:BL4}
\end{table}

\begin{table}[t]
\hspace*{-1cm}
\begin{tabular}{cccccccc}
\hline \hline
 \# & Decompositions & \multicolumn{2}{c}{Mediators} 
 & \begin{minipage}{3cm}
    Projection to \\
    the basis ops.
    \end{minipage}
 & $m_{\nu}$@tree
 & $m_{\nu}$@1loop
 & $m_{\nu}$@2loop
 \\
 \hline
 \#1 
 & $(L_{\alpha}e_{R \beta})(H)(\overline{d_{R}}u_{R})$
     & $V({\bf 1}, {\bf 2})_{+\frac{3}{2}}$
	 & $V'({\bf 1}, {\bf 1})_{-1}$
	     & $\mathcal{O}_{8}(\alpha,\beta)$
		 & ---
		     &  
 \begin{minipage}{2.1cm}
  \vspace{0.1cm}
  $\left[
  \begin{minipage}{2cm}
   T$\nu$I-i
  \\
  w. $\partial HHV'$
  \end{minipage}
   \right]$
 \end{minipage}
 &
  \begin{minipage}{1.6cm}
   \vspace{0.1cm}
     $2 \times T2^{\rm B}_{4}$ 
 \end{minipage}
 		 \\
 \hline
 \#2 
 & $(L_{\alpha} u_{R})(H)(\overline{d_{R}}e_{R \beta})$
     & $V(\overline{\bf 3}, {\bf 2})_{-\frac{1}{6}}$
	 & $V'(\overline{\bf 3}, {\bf 1})_{-\frac{2}{3}}$
	     & $\mathcal{O}_{8}(\alpha,\beta)$
		 & ---
		     & 
 \begin{minipage}{2.1cm}
  \vspace{0.1cm}
  T$\nu$I-ii
  \\
  w. $\overline{Q}L V'^{\dagger}$
  \vspace{0.1cm}
 \end{minipage}
			 &
 $T2^{\rm B}_{2}$+$T2^{\rm B}_{4}$
 \\
 \hline
 \#3 
 & $(\overline{d_{R}} L_{\alpha})(H)(u_{R} e_{R \beta})$
     & $S({\bf 3}, {\bf 2})_{+\frac{1}{6}}$
	 & $S'({\bf 3}, {\bf 1})_{-\frac{1}{3}}$
	     & $\frac{1}{2}\mathcal{O}_{8}(\alpha,\beta)$
		 & ---
		     &
 \begin{minipage}{2.1cm}
  \vspace{0.1cm}
  T$\nu$I-ii
  \\
  w. $QL S'^{\dagger}$
  \vspace{0.1cm}
 \end{minipage}
			 &
  \begin{minipage}{2cm}
   \vspace{0.1cm}
   $T2^{\rm B}_{2}$+$T2^{\rm B}_{4}$
   \\
   $\mathcal{O}_{8}^{4}$ in \cite{Cai:2014kra},
   \cite{Babu:2010vp}
 \end{minipage}
 \\
 \hline
 \#4 
 & $(L_{\alpha} e_{R\beta})(u_{R})(\overline{d_{R}} H)$
     & $V({\bf 1}, {\bf 2})_{+\frac{3}{2}}$
	 & $\psi_{L,R}({\bf 3}, {\bf 2})_{-\frac{5}{6}}$
	     & $\mathcal{O}_{8}(\alpha,\beta)$
		 & ---
		     & ---
			 &
$T2^{\rm B}_{1}$+$T2^{\rm B}_{2}$
 \\
 \hline
 \#5 
 & $(L_{\alpha} H)(u_{R})(\overline{d_{R}} e_{R\beta})$
     & $\psi_{R}({\bf 1}, {\bf 1})_{0}$
	 & $V(\overline{\bf 3},{\bf 1})_{-\frac{2}{3}}$
	     & $\mathcal{O}_{8}(\alpha,\beta)$
		 & 
		      type I
 & 
     & 
 \\
 \hline
 \#6 
 & $(\overline{d_{R}} L_{\alpha})(u_{R})(e_{R\beta} H)$
     & $S({\bf 3}, {\bf 2})_{+\frac{1}{6}}$
	 & $\psi_{L,R}({\bf 1},{\bf 2})_{-\frac{1}{2}}$
	     & $\frac{1}{2}\mathcal{O}_{8}(\alpha,\beta)$
		 & 
		     ---
		     & ---
			 &
  \begin{minipage}{2cm}
    \vspace{0.1cm}
   $T2^{\rm B}_{1}$+$T2^{\rm B}_{2}$
  \\
   $\mathcal{O}_{8}^{2}$ in \cite{Cai:2014kra}
 \end{minipage}
 \\ 
 \hline
 \#7 
 & $(L_{\alpha} e_{R \beta})(\overline{d_{R}})(u_{R} H)$
     & $V({\bf 1}, {\bf 2})_{+\frac{3}{2}}$
	 & $\psi_{L,R} ({\bf 3}, {\bf 2})_{+\frac{7}{6}}$
	     & $\mathcal{O}_{8}(\alpha,\beta)$
		 & 
		     ---
		     & ---
			 &
$T2^{\rm B}_{1}$+$T2^{\rm B}_{2}$
 \\ 
 \hline
 \#8 
 & $(L_{\alpha} H)(\overline{d_{R}})(u_{R} e_{R\beta})$
     & $\psi_{R}({\bf 1}, {\bf 1})_{0}$
	 & $S ({\bf 3}, {\bf 1})_{-\frac{1}{3}}$
	     & $\frac{1}{2} \mathcal{O}_{8}(\alpha,\beta)$
		 &
		      type I
 & 
     & 
 \\ 
 \hline
 \#9 
 & $(L_{\alpha} u_{R})(\overline{d_{R}})(e_{R\beta} H)$
     & $V(\overline{\bf 3}, {\bf 2})_{-\frac{1}{6}}$
	 & $\psi_{L,R} ({\bf 1}, {\bf 2})_{-\frac{1}{2}}$
	     & $\mathcal{O}_{8}(\alpha,\beta)$
		 & ---
		     & ---
			 &   
$T2^{\rm B}_{1}$+$T2^{\rm B}_{2}$
 \\ 
 \hline
 \#10 
 & $(e_{R\beta} H)(L_{\alpha})(\overline{d_{R}} u_{R})$
     & $\psi_{L,R}({\bf 1}, {\bf 2})_{+\frac{1}{2}}$
	 & $V ({\bf 1}, {\bf 1})_{+1}$
	     & $-\mathcal{O}_{8}(\alpha,\beta)$
		 & ---
 & 			 \begin{minipage}{2.1cm}
			 \vspace{0.1cm}
			  $\left[
			  \begin{minipage}{2cm}
			   T$\nu$I-ii\\
			   w. $\partial HHV$ 
			  \end{minipage}
			  \right]$
			 \end{minipage}
			 &
$2\times T2^{\rm B}_{4}$
 \\ 
 \hline
 \#11 
 & $(e_{R\beta} u_{R})(L_{\alpha})(\overline{d_{R}} H)$
     & $S(\overline{\bf 3}, {\bf 1})_{+\frac{1}{3}}$
	 & $\psi_{L,R} ({\bf 3}, {\bf 2})_{-\frac{5}{6}}$
	     & $-\frac{1}{2}\mathcal{O}_{8}(\alpha,\beta)$
		 & ---
		     & 
  \begin{minipage}{2.1cm}
  \vspace{0.1cm}
  T$\nu$I-iii
  \\
  w. $QL S$
  \vspace{0.1cm}
 \end{minipage}
			 &   
 \begin{minipage}{2cm}
  \vspace{0.1cm}
  $T2^{\rm B}_{2}$+$T2^{\rm B}_{4}$
  \\
  $\mathcal{O}_{8}^{1}$ in \cite{Cai:2014kra}
 \end{minipage}
 \\ 
 \hline
 \#12 
 & $(\overline{d_{R}} e_{R\beta})(L_{\alpha})(u_{R} H)$
     & $V({\bf 3}, {\bf 1})_{+\frac{2}{3}}$
	 & $\psi_{L,R} ({\bf 3}, {\bf 2})_{+\frac{7}{6}}$
	     & $\mathcal{O}_{8}(\alpha,\beta)$
		 & ---
		     & 
 \begin{minipage}{2.1cm}
  \vspace{0.1cm}
  T$\nu$I-iii
  \\
  w. $\overline{Q}L V$
  \vspace{0.1cm}
 \end{minipage}
 &   
$T2^{\rm B}_{2}$+$T2^{\rm B}_{4}$
 \\ 
 \hline
 \#13 
 & $(L_{\alpha} u_{R})(e_{R\beta})(\overline{d_{R}} H)$
     & $V(\overline{\bf 3}, {\bf 2})_{-\frac{1}{6}}$
	 & $\psi_{L,R} ({\bf 3}, {\bf 2})_{-\frac{5}{6}}$
	     & $-\mathcal{O}_{8}(\alpha,\beta)$
		 & ---
		     & ---
			 &   
$2 \times T2^{\rm B}_{2}$
 \\ 
 \hline
 \#14 
 & $(L_{\alpha} H)(e_{R\beta})(\overline{d_{R}} u_{R})$
     & $\psi_{R}({\bf 1}, {\bf 1})_{0}$
	 & $V ({\bf 1}, {\bf 1})_{+1}$
	     & $\mathcal{O}_{8}(\alpha,\beta)$
		 & 
		    type I
		     & 
			 & 
 \\ 
 \hline
 \#15 
 & $(\overline{d_{R}} L_{\alpha})(e_{R\beta})(u_{R} H)$
     & $S(\overline{\bf 3}, {\bf 2})_{+\frac{1}{6}}$
	 & $\psi_{L,R} ({\bf 3}, {\bf 2})_{+\frac{7}{6}}$
	     & $\frac{1}{2}\mathcal{O}_{8}(\alpha,\beta)$
		 & ---
		     & ---
			 &   
 \begin{minipage}{2cm}
  \vspace{0.1cm}
$2\times T2^{\rm B}_{2}$
  \\
  $\mathcal{O}_{8}^{3}$ in \cite{Cai:2014kra}
 \end{minipage}
 \\ 
 \hline \hline
\end{tabular}
\caption{Decomposition of the $Le_{R} u_{R} \overline{d_{R}} H$
 operator. The long-range contribution \cite{Doi:1985dx} 
 in left-right symmetric models
 \cite{Pati:1974yy,Mohapatra:1974gc,Mohapatra:1980yp}
 corresponds to decomposition \#14~.}
\label{Tab:BL8}
\end{table}


There is another category of lepton-number-violating 
effective operators, not contained in the catalogue by Babu and Leung: 
operators with covariant derivatives $D_{\rho}$. 
These have been intensively studied 
in Refs.~\cite{delAguila:2012nu,Lehman:2014jma,Bhattacharya:2015vja}.
The derivative operators with mass dimension seven are 
classified into two types by their ingredient fields;
One is $D_{\rho} D^{\rho} LL HH$ and the other 
is $D_{\rho} L \gamma^{\rho} e_{R} HHH$.
With the full decomposition,
it is straightforward 
to show that the tree-level decompositions of the first type
must contain one of the seesaw mediators.
Therefore, the neutrino masses are generated at the tree level
and the mass mechanism always dominate the contributions 
to $\znbb$.
The decompositions of the second type also require
the scalar triplet of the type II seesaw mechanism
when we do not employ vector fields as mediators,
and the new physics contributions to $\znbb$ 
become insignificant again compared to the mass mechanism.
In Ref.~\cite{delAguila:2012nu},
the authors successfully
obtained the derivative operator 
$
(\overline{{e_{R}}^{c}} \gamma^{\rho} 
L {\rm i} \tau^{2} \vec{\tau} \vec{W}_{\rho} H')
(H{\rm i} \tau^{2} H')
$
at the tree level and simultaneously
avoided the tree-level neutrino mass
with the help of a second Higgs doublet $H'({\bf 1},{\bf 2})_{+1/2}$  
and a $Z_{2}$ parity which is broken spontaneously.
Here we restrict ourselves to use the ingredients obtained 
from decompositions and do not discuss such extensions.
Within our framework, the derivative operators are 
always associated with tree-level neutrino masses.
In this study, we have mainly focused on the cases
where the new physics contributions give
a considerable impact on the $\znbb$ processes.
Therefore, we do not go into the details of
the decompositions of the derivative operators.


\begin{thebibliography}{10}

\bibitem{Deppisch:2012nb}
F.~F. Deppisch, M.~Hirsch, and H.~P\"as,
\newblock J.Phys. {\bf G39}, 124007 (2012), arXiv:1208.0727.

\bibitem{Hirsch:2015cga}
M.~Hirsch,
\newblock AIP Conf. Proc. {\bf 1666}, 170007 (2015).

\bibitem{Schechter:1981bd}
J.~Schechter and J.~Valle,
\newblock Phys.Rev. {\bf D25}, 2951 (1982).

\bibitem{Duerr:2011zd}
M.~Duerr, M.~Lindner, and A.~Merle,
\newblock JHEP {\bf 1106}, 091 (2011), arXiv:1105.0901.

\bibitem{Forero:2014bxa}
D.~Forero, M.~Tortola, and J.~Valle,
\newblock Phys.Rev. {\bf D90}, 093006 (2014), arXiv:1405.7540.

\bibitem{Deppisch:2013jxa}
F.~F. Deppisch, J.~Harz, and M.~Hirsch,
\newblock Phys.Rev.Lett. {\bf 112}, 221601 (2014), arXiv:1312.4447.

\bibitem{Deppisch:2015yqa}
F.~F. Deppisch, J.~Harz, M.~Hirsch, W.-C. Huang, and H.~P\"as,
\newblock (2015), arXiv:1503.04825.

\bibitem{Weinberg:1979sa}
S.~Weinberg,
\newblock Phys. Rev. Lett. {\bf 43}, 1566 (1979).

\bibitem{Yanagida:1979as}
T.~Yanagida,
\newblock Conf.Proc. {\bf C7902131}, 95 (1979).

\bibitem{GellMann:1980vs}
M.~Gell-Mann, P.~Ramond, and R.~Slansky,
\newblock Conf.Proc. {\bf C790927}, 315 (1979),
\newblock Supergravity, P. van Nieuwenhuizen and D.Z. Freedman (eds.), North
  Holland Publ. Co., 1979.

\bibitem{Foot:1988aq}
R.~Foot, H.~Lew, X.~He, and G.~C. Joshi,
\newblock Z.Phys. {\bf C44}, 441 (1989).

\bibitem{Mohapatra:1979ia}
R.~N. Mohapatra and G.~Senjanovic,
\newblock Phys. Rev. Lett. {\bf 44}, 912 (1980).

\bibitem{Ma:1998dn}
E.~Ma,
\newblock Phys.Rev.Lett. {\bf 81}, 1171 (1998), arXiv:hep-ph/9805219.

\bibitem{Babu:2001ex}
K.~Babu and C.~N. Leung,
\newblock Nucl.Phys. {\bf B619}, 667 (2001), arXiv:hep-ph/0106054.

\bibitem{Giudice:2008uua}
G.~F. Giudice and O.~Lebedev,
\newblock Phys. Lett. {\bf B665}, 79 (2008), arXiv:0804.1753.

\bibitem{Babu:2009aq}
K.~S. Babu, S.~Nandi, and Z.~Tavartkiladze,
\newblock Phys. Rev. {\bf D80}, 071702 (2009), arXiv:0905.2710.

\bibitem{Bonnet:2009ej}
F.~Bonnet, D.~Hernandez, T.~Ota, and W.~Winter,
\newblock JHEP {\bf 0910}, 076 (2009), arXiv:0907.3143.

\bibitem{Picek:2009is}
I.~Picek and B.~Radovcic,
\newblock Phys. Lett. {\bf B687}, 338 (2010), arXiv:0911.1374.

\bibitem{Kanemura:2010bq}
S.~Kanemura and T.~Ota,
\newblock Phys. Lett. {\bf B694}, 233 (2011), arXiv:1009.3845.

\bibitem{Krauss:2011ur}
M.~B. Krauss, T.~Ota, W.~Porod, and W.~Winter,
\newblock Phys. Rev. {\bf D84}, 115023 (2011), arXiv:1109.4636.

\bibitem{Krauss:2013gy}
M.~B. Krauss, D.~Meloni, W.~Porod, and W.~Winter,
\newblock JHEP {\bf 05}, 121 (2013), arXiv:1301.4221.

\bibitem{Bambhaniya:2013yca}
G.~Bambhaniya, J.~Chakrabortty, S.~Goswami, and P.~Konar,
\newblock Phys. Rev. {\bf D88}, 075006 (2013), arXiv:1305.2795.

\bibitem{Pas:1999fc}
H.~P\"as, M.~Hirsch, H.~Klapdor-Kleingrothaus, and S.~Kovalenko,
\newblock Phys.Lett. {\bf B453}, 194 (1999).

\bibitem{Pas:2000vn}
H.~P\"as, M.~Hirsch, H.~Klapdor-Kleingrothaus, and S.~Kovalenko,
\newblock Phys.Lett. {\bf B498}, 35 (2001), arXiv:hep-ph/0008182.

\bibitem{Bonnet:2012kh}
F.~Bonnet, M.~Hirsch, T.~Ota, and W.~Winter,
\newblock JHEP {\bf 1303}, 055 (2013), arXiv:1212.3045.

\bibitem{Helo:2015fba}
J.~Helo, M.~Hirsch, T.~Ota, and F.~A.~P. Dos~Santos,
\newblock JHEP {\bf 1505}, 092 (2015), arXiv:1502.05188.

\bibitem{deGouvea:2007xp}
A.~de~Gouvea and J.~Jenkins,
\newblock Phys.Rev. {\bf D77}, 013008 (2008), arXiv:0708.1344.

\bibitem{Angel:2012ug}
P.~W. Angel, N.~L. Rodd, and R.~R. Volkas,
\newblock Phys.Rev. {\bf D87}, 073007 (2013), arXiv:1212.6111.

\bibitem{Angel:2013hla}
P.~W. Angel, Y.~Cai, N.~L. Rodd, M.~A. Schmidt, and R.~R. Volkas,
\newblock JHEP {\bf 1310}, 118 (2013), arXiv:1308.0463.

\bibitem{Helo:2013ika}
J.~Helo, M.~Hirsch, H.~P\"as, and S.~Kovalenko,
\newblock Phys.Rev. {\bf D88}, 073011 (2013), arXiv:1307.4849.

\bibitem{Helo:2015ffa}
J.~C. Helo and M.~Hirsch,
\newblock Phys. Rev. {\bf D92}, 073017 (2015), arXiv:1509.00423.

\bibitem{Babu:1995vh}
K.~Babu and R.~Mohapatra,
\newblock Phys.Rev.Lett. {\bf 75}, 2276 (1995), arXiv:hep-ph/9506354.

\bibitem{Pas:1998nn}
H.~P\"as, M.~Hirsch, and H.~Klapdor-Kleingrothaus,
\newblock Phys.Lett. {\bf B459}, 450 (1999), arXiv:hep-ph/9810382.

\bibitem{Hirsch:1996ye}
M.~Hirsch, H.~Klapdor-Kleingrothaus, and S.~Kovalenko,
\newblock Phys.Rev. {\bf D54}, 4207 (1996), arXiv:hep-ph/9603213.

\bibitem{Pati:1974yy}
J.~C. Pati and A.~Salam,
\newblock Phys.Rev. {\bf D10}, 275 (1974).

\bibitem{Mohapatra:1974gc}
R.~Mohapatra and J.~C. Pati,
\newblock Phys.Rev. {\bf D11}, 2558 (1975).

\bibitem{Mohapatra:1980yp}
R.~N. Mohapatra and G.~Senjanovic,
\newblock Phys. Rev. {\bf D23}, 165 (1981).

\bibitem{Cai:2014kra}
Y.~Cai, J.~D. Clarke, M.~A. Schmidt, and R.~R. Volkas,
\newblock JHEP {\bf 1502}, 161 (2015), arXiv:1410.0689.

\bibitem{Lehman:2014jma}
L.~Lehman,
\newblock Phys. Rev. {\bf D90}, 125023 (2014), arXiv:1410.4193.

\bibitem{Bhattacharya:2015vja}
S.~Bhattacharya and J.~Wudka,
\newblock (2015), arXiv:1505.05264.

\bibitem{Gando:2012zm}
KamLAND-Zen Collaboration, A.~Gando {\em et~al.},
\newblock Phys. Rev. Lett. {\bf 110}, 062502 (2013), arXiv:1211.3863.

\bibitem{Koide:2001xy}
Y.~Koide,
\newblock Phys. Rev. {\bf D64}, 077301 (2001), arXiv:hep-ph/0104226.

\bibitem{Babu:2011vb}
K.~Babu and J.~Julio,
\newblock Phys.Rev. {\bf D85}, 073005 (2012), arXiv:1112.5452.

\bibitem{Haxton:1985am} 
  W.~C.~Haxton and G.~J.~Stephenson,
  Prog.\ Part.\ Nucl.\ Phys.\  {\bf 12}, 409 (1984).

\bibitem{Muto:1989cd} 
  K.~Muto, E.~Bender and H.~V.~Klapdor,
  Z.\ Phys.\ A {\bf 334}, 187 (1989).


\bibitem{Antusch:2006vwa}
S.~Antusch, C.~Biggio, E.~Fernandez-Martinez, M.~B. Gavela, and J.~Lopez-Pavon,
\newblock JHEP {\bf 10}, 084 (2006), arXiv:hep-ph/0607020.

\bibitem{Akhmedov:2013hec}
E.~Akhmedov, A.~Kartavtsev, M.~Lindner, L.~Michaels, and J.~Smirnov,
\newblock JHEP {\bf 05}, 081 (2013), arXiv:1302.1872.

\bibitem{Antusch:2014woa}
S.~Antusch and O.~Fischer,
\newblock JHEP {\bf 10}, 94 (2014), arXiv:1407.6607.

\bibitem{Escrihuela:2015wra}
F.~J. Escrihuela, D.~V. Forero, O.~G. Miranda, M.~Tortola, and J.~W.~F. Valle,
\newblock Phys. Rev. {\bf D92}, 053009 (2015), arXiv:1503.08879.

\bibitem{delAguila:2012nu}
F.~del Aguila, A.~Aparici, S.~Bhattacharya, A.~Santamaria, and J.~Wudka,
\newblock JHEP {\bf 06}, 146 (2012), arXiv:1204.5986.

\bibitem{Bonnet:2012kz}
F.~Bonnet, M.~Hirsch, T.~Ota, and W.~Winter,
\newblock JHEP {\bf 1207}, 153 (2012), arXiv:1204.5862.

\bibitem{Hirsch:1996qy}
M.~Hirsch, H.~Klapdor-Kleingrothaus, and S.~Kovalenko,
\newblock Phys.Lett. {\bf B378}, 17 (1996), arXiv:hep-ph/9602305.

\bibitem{AristizabalSierra:2007nf}
D.~Aristizabal~Sierra, M.~Hirsch, and S.~Kovalenko,
\newblock Phys.Rev. {\bf D77}, 055011 (2008), arXiv:0710.5699.

\bibitem{Stupak:2012aj}
ATLAS, J.~Stupak~III,
\newblock EPJ Web Conf. {\bf 28}, 12012 (2012), arXiv:1202.1369.

\bibitem{ATLAS:2012aq}
ATLAS, G.~Aad {\em et~al.},
\newblock Eur.Phys.J. {\bf C72}, 2151 (2012), arXiv:1203.3172.

\bibitem{CMS:2014qpa}
CMS, C.~Collaboration,
\newblock (2014), CMS-PAS-EXO-12-041.

\bibitem{Khachatryan:2015bsa}
CMS, V.~Khachatryan {\em et~al.},
\newblock (2015), arXiv:1503.09049.

\bibitem{Khachatryan:2014ura}
CMS, V.~Khachatryan {\em et~al.},
\newblock Phys.Lett. {\bf B739}, 229 (2014), arXiv:1408.0806.

\bibitem{Sierra:2014rxa}
D.~Aristizabal~Sierra, A.~Degee, L.~Dorame, and M.~Hirsch,
\newblock JHEP {\bf 1503}, 040 (2015), arXiv:1411.7038.

\bibitem{Lavoura:2003xp}
L.~Lavoura,
\newblock Eur.Phys.J. {\bf C29}, 191 (2003), arXiv:hep-ph/0302221.

\bibitem{Arnold:2010tu}
SuperNEMO Collaboration, R.~Arnold {\em et~al.},
\newblock Eur.Phys.J. {\bf C70}, 927 (2010), arXiv:1005.1241.

\bibitem{Casas:2001sr}
J.~Casas and A.~Ibarra,
\newblock Nucl.Phys. {\bf B618}, 171 (2001), arXiv:hep-ph/0103065.

\bibitem{Adam:2013mnn}
MEG Collaboration, J.~Adam {\em et~al.},
\newblock Phys.Rev.Lett. {\bf 110}, 201801 (2013), arXiv:1303.0754.

\bibitem{Doi:1985dx}
M.~Doi, T.~Kotani, and E.~Takasugi,
\newblock Prog.Theor.Phys.Suppl. {\bf 83}, 1 (1985).

\bibitem{Beringer:1900zz}
Particle Data Group, J.~Beringer {\em et~al.},
\newblock Phys.Rev. {\bf D86}, 010001 (2012).

\bibitem{Khachatryan:2014dka}
CMS, V.~Khachatryan {\em et~al.},
\newblock Eur. Phys. J. {\bf C74}, 3149 (2014), arXiv:1407.3683.

\bibitem{Aad:2015xaa}
ATLAS, G.~Aad {\em et~al.},
\newblock JHEP {\bf 07}, 162 (2015), arXiv:1506.06020.

\bibitem{Babu:2010vp}
K.~S. Babu and J.~Julio,
\newblock Nucl. Phys. {\bf B841}, 130 (2010), arXiv:1006.1092.

\end{thebibliography}
\end{document}